\documentclass[aps,prd,11pt, preprintnumbers,numbers,sort&compress,nofootinbib]{revtex4}
\pdfoutput=1

\usepackage{amsmath}
\usepackage{graphicx}
\usepackage{amssymb}
\usepackage{hyperref}

\newcommand{\nn}{\nonumber \\}
\newcommand{\n}{\nabla}
\newcommand{\rref}[1]{(\ref{#1})}

\newcommand{\exclude}[1]{}

\newcommand{\beq}{\begin{equation}}
\newcommand{\eeq}{\end{equation}}
\newcommand{\bea}{\begin{eqnarray}}
\newcommand{\eea}{\end{eqnarray}}

\newcommand{\q}{{\bf q}}

\newcommand{\p}{\partial}

\newcommand{\OO}{{\cal O}}
\newcommand{\w}{{\bf w}}
\newcommand{\NN}{{\cal N}}

\newcommand{\junk}[1]{}

\begin{document}

\title{Fluctuations in finite density holographic quantum liquids} 
  \author{ Mikhail Goykhman, Andrei Parnachev and Jan Zaanen}
   \affiliation {Institute Lorentz for Theoretical Physics, Leiden University, P.O. Box 9506, Leiden 2300RA, The Netherlands}


\begin{abstract}
We study correlators of the global $U(1)$ currents in holographic models which involve $\NN=4$ SYM coupled
to the finite density matter in the probe brane sector.
We find the spectral density associated with the longitudinal response to be exhausted by the zero sound pole 
and argue that this could be consistent with the behavior of Fermi liquid with vanishing Fermi velocity.
However the transversal response shows an unusual momentum independent behavior.
Inclusion of magnetic field leads to a gap in the dispersion relation for the zero sound mode propagating in the plane
of magnetic field.
For small values of the magnetic field $B$ the gap in the spectrum scales linearly with $B$,
which is consistent with Kohn's theorem for nonrelativistic fermions with pairwise interaction.
We do not find signatures of multiple Landau levels expected in Landau Fermi liquid theory.
We also consider the influence of generic higher derivative corrections on the form of the spectral function.
\end{abstract}
 

\maketitle

\section{Introduction and summary}

Perhaps the deepest open problem in condensed matter physics is the classification of
compressible quantum liquids. This refers to stable states of zero temperature quantum matter that do not 
break any symmetry and support massless excitations. 
This question cannot be easily addressed within the confines of standard field theory. 
The issue arises when fermions are considered at finite density and the culprit is known as
the ``fermion sign" problem.
Dealing with time-reversal symmetric finite density bosonic matter the methods of equilibrium statistical physics 
give a full  control and invariably one finds that the ground states break symmetry. Dealing with 
incompressible quantum fluids like the fractional quantum Hall states the mass gap is quite instrumental to 
control the theory, revealing the profound non-classical phenomenon of topological order. The hardship is with the 
compressible quantum fluids: the only example which is fully understood is the Fermi-liquid.       

The ease of the mathematical description of the Fermi-liquid as the adiabatic continuation of the Fermi-gas is in a way
deceptive. Compared to classical fluids its low energy spectrum of non-charged excitations 
is amazingly rich. In addition to the  zero sound, there is a continuum
of  volume conserving  ``shape fluctuations"  of the Fermi-surface, corresponding with  the particle-hole excitations
(Lindhard continuum) of the conventional perturbative lore. Although serious doubts exist regarding the mathematical consistency
and their relevance towards real physics, the ``fractionalized (spin) liquids" that were constructed in condensed matter physics 
appear to be still controlled by the presence of a  Fermi-surface while these are not Landau Fermi-liquids in the strict sense. 
This inspired Sachdev to put forward the interesting conjecture that the Fermi-surface might be ubiquitous for all compressible quantum liquids \cite{Sachdev:2011wg}. 

The gauge-gravity duality or AdS/CFT correspondence provides a unique framework to deal with these matters in a 
controlled way
(see \cite{Sachdev:2011wg, arXiv:0909.0518,arXiv:0904.1975,arXiv:1002.1722,Hartnoll:2011fn,Iqbal:2011ae} for recent reviews).
Although it addresses field theories that are at first sight very remote from the interacting electrons of condensed matter, there are reasons to believe
that it reveals generic emergence phenomena associated with strongly interacting quantum systems. 
Field theories whose understanding is plagued by the ``fermion sign" problem appear to be quite tractable
in the dual gravitational description. 
With regard to unconventional Fermion  physics, perhaps the most important achievement has been the discovery of the ``AdS$_2$ 
metal" \cite{arXiv:0903.2477,Cubrovic:2009ye}, dual to the asymptotically AdS  Reissner-Nordstrom black hole. 
On the field theory side this describes a local (purely temporal) quantum critical state that was not  expected on basis of conventional
 field theoretic means. 
Although quite promising regarding the intermediate 
temperature physics (the ``strange" normal states) in high Tc superconductors and so forth, this AdS$_2$ metal is probably
 not a stable state, given its zero
temperature entropy.  Much of recent activity has been devoted to the study of the instability of this metal towards bosonic  symmetry breaking (holographic 
superconductivity   \cite{pair}, ``stripe" instabilities \cite{Donos:2011qt})
and towards the stable Fermi liquid \cite{Hartnoll:2010gu, Cubrovic:2010bf, Sachdev:2011ze}.

The  top-down constructions might become quite instrumental in facilitating the search for truly 
new quantum liquids. An important category are the Dp/Dq brane intersections; the $p=3$ case provides us 
with a set of especially tractable examples. 
The dynamics of the low energy degrees of freedom of the D3-Dp strings
can be studied in the probe approximation where the back-reaction to the 
$AdS_5\times S^5$ geometry can be neglected \cite{Karch:2002sh}.
In this paper we will consider D3 and Dp branes intersecting along 2+1 dimensions, where 
p=5 (p=7) corresponds to the (non)-supersymmetric system.
As emphasized in \cite{Davis:2011gi} the nonsupersymmetric system can be viewed as a model of graphene: the brane intersection fermions are 
like the Dirac fermions moving on the 2+1D graphene backbone,
(tunable to finite density by gating), interacting strongly through the gauge fields living in 3+1 dimensions.
We will present a number of results for the longitudinal- and transversal {\em dynamical} charge susceptibilities (at finite frequency $\w$ and 
momenta $\q$)\footnote{In this paper we denote the values of frequency and momentum by bold letters.
The usual letters, defined below, are reserved for dimensionless variables.}, in the
absence and presence of a magnetic field, for both the supersymmetric and non-supersymmetric D3/Dp systems at finite density. 
We find very similar results in both the supersymmetric- and fermionic set ups, showing that these outcomes at strong 't Hooft coupling
are not caused by the difference in the Lagrangians.  We find suggestive
indications for the presence of an entirely new form of quantum liquid, but we cannot be entirely conclusive. 
Our observations cannot entirely rule out the existence of a Fermi liquid with vanishing Fermi velocity.  

In fact, the first study of these systems at finite density already produced evidence that some odd state is created. 
In ref.  \cite{Karch:2008fa}  it was observed that the density-dependent part of the heat capacity in the D3/Dp systems
with $2+1$ dimensional intersection
behaves like $T^4$.
This is in contrast to the result for the 
Fermi-liquids which is set by the Sommerfeld law of the specific heat $C = \gamma T$, where the Sommerfeld coefficient $\gamma$ is proportional to the quasiparticle mass. 
This behavior remains to be understood: for example, it is conceivable that the linear term in the
heat capacity exists, but is parametrically suppressed in the holographic model.
On a side, it is worth noting that in the context of pnictide 
superconductivity  a rogue signal has been detected that refuses to disappear: this indicates that the electronic specific heat of the metal state $\sim T^3$ 
\cite{pnictide}.
      
As mentioned above, besides the Lindhard continuum an interacting Fermi liquid will carry a single propagating mode called zero sound. 
Unlike the usual sound at finite temperature, translational invariance alone is not sufficient for establishing the
existence of the zero sound mode. 
 The discovery 
of zero sound associated with the brane intersection matter \cite{Karch:2008fa} is therefore significant. 
The fate of the holographic  zero sound was further studied in 
\cite{Kulaxizi:2008kv,Kulaxizi:2008jx,arXiv:1005.4075,arXiv:1007.0590,arXiv:1009.3966,Bergman:2011rf,Davison:2011ek,arXiv:1111.0660}
(see also  \cite{arXiv:1003.1134,arXiv:1108.1798} for closely related work).  
At very low temperature the attenuation (damping) of this zero sound behaves
like the (``collisionless") Fermi liquid zero sound, in the sense that it increases like the square of its momentum. 
In \cite{Davison:2011ek} it was found also that upon increasing temperature the zero sound 
velocity decreases while the attenuation increases, turning into a purely diffusive pole at high temperatures. This is different from the crossover 
from zero sound to ordinary sound as function of temperature in a single component Fermi-system like $^3He$. 
In the brane intersection systems momentum 
is shared between the superconformal strongly coupled uncharged sector
 and the material system on the intersection, and the latter does not support hydrodynamical sound
in isolation. Somehow, upon lowering temperature the momentum of the brane intersection matter becomes separately conserved, facilitating the emergence of the 
zero sound in the low temperature limit. 


Given that zero sound is rather ubiquitous, one would like to obtain more direct information regarding the density fluctuations of the quantum liquid. These are
expected to be contained in the fully dynamical, momentum and energy dependent charge susceptibility/density-density propagator associated with the conserved
charge on the brane-intersection. 
One strategy is to look for the momentum dependence of the reactive response (real part) at zero frequency: one expects a singularity
at twice the Fermi momentum,
$2 \q_F$ where the Luttinger's theorem implies that $\q_F$ is set by
the bare chemical potential, $\q_F \sim \mu$ .
A number of papers has been devoted to the search of such singular behavior in the framework of AdS/CFT.
In \cite{Kulaxizi:2008jx} the $\langle J^0 J^0\rangle$ correlator has been computed in the holographic
setup where the only charged degrees of freedom are four-dimensional fermions.
The resulting function was completely smooth.
In \cite{Gauntlett:2011mf,arXiv:1106.6030,arXiv:1108.1205,arXiv:1110.4601} the two-point
function for global currents was computed for various systems and again the tree-level computation
in the bulk did not show any nonanalytic behavior.
Very recently it has been argued that a singularity can be observed in the systems where an exact
result to all orders in  $\alpha'$ is available \cite{Polchinski:2012nh}.

Searching for the singularity at $2 \q_F$ is in  principle a tricky procedure because these ``Friedel oscillation" singularities are strongly weakened
by the self energy effects in the strongly interacting Fermi-liquid. Another way to probe for the signatures of the
Fermi liquid   is to compute the imaginary part
of the dynamical density susceptibility in a large kinematical window because this 
spectral function shows directly the density excitations of the system. The result is well known in the weakly interacting Fermi liquid, see Fig. 1: besides the zero sound 
pole one finds the Lindhard continuum of particle hole excitations. 
It is worth noting that as the value of the Landau parameter $F_0$  increases, the spectral weight in the density response  is increasingly concentrated in the zero sound poles, ``hiding" the Lindhard 
continuum. In this regard the {\em transversal} density propagator is quite informative: since in this channel no collective modes are expected to form,
 this is the place to look for the incoherent Fermi-surface fluctuations.
Unfortunately technical issues prevent us from accessing the regime of parametrically small 
Fermi velocity.
Our holographic computations of the longitudinal- and transversal dynamical charge susceptibilities are limited to a kinematical window 
where $\w\sim|\q| $. 

Despite this caveat, the holographic density propagators that we compute reveal very interesting information. We find that the longitudinal density propagator 
is within our numerical resolution completely exhausted by the zero sound pole (Fig. 4). Regardless the precise nature of the underlying state this signals  
very strong density/density correlations in this liquid. The transversal charge propagator shows that sound is not the whole story. The ``other stuff", albeit very unlike 
a Lindhard continuum, signals the presence of a sector of highly collective, deep IR density fluctuations: the imaginary part of the 
transversal propagator behaves like $\chi^{(i)}_t (\q, \w) \sim \w$.
This response is surprisingly momentum independent and suggests local quantum criticality, which
was instrumental in the "AdS$_2$ metal" setup.
All of this seems to imply that we are indeed dealing with some entirely new quantum liquid.

To probe some of the features of this  quantum liquid, we introduce an external magnetic field 
which is a valuable ``experimental tool".
 This induces the  gap in the spectrum  that is visible in the holographic calculations.
 Dealing with a 2+1D Fermi-liquid one would expect the signatures of Landau levels also in the density response. 
 In the strongly interacting system, the longitudinal response should reveal the ``magneto-roton",
the left over of zero sound in the system with a magnetic field which is well known from (fractional) quantum Hall systems \cite{GMP}
\footnote{See \cite{Jokela:2010nu} for related work in the context of holography.}.
According to Kohn's theorem \cite{Kohn:1961zz}, the density spectrum should show a gap equal to the cyclotron frequency at zero momentum.
Note that this theorem is very generic and only assumes that  degrees of freedom, charged under the magnetic field, interact  pairwise.
Our holographic calculation reveals that:
i) at small values of the magnetic field $B$ the value of the gap\footnote{This is also consistent
with the observations made in \cite{Bergman:2011rf,Jokela:2012vn} where the same D3/D7 system,
modified by the inclusion of flux through the internal cycles, is considered.}
scales linearly with $B$, which
is  consistent with  Kohn's theorem for the nonrelativistic fermions and
ii) there are no signatures of Landau levels associated with incoherent particle-hole excitations (Fig. 2).



The remainder of this paper is organized as follows.
The next section is devoted to the review of Landau Fermi liquid theory 
including the random phase approximation (RPA) for the dynamical response.
In particular, we review the appearance of the zero sound mode in the RPA calculation
of the density-density correlator.
As the value of the interaction strength increases, the Lindhard continuum gets separated
from the zero sound pole (Fig. 1) and gradually disappears. 
In the extreme limit of vanishing Fermi velocity, the spectral density
is completely exhausted by the zero sound mode.
We also review the RPA expectations for the 2+1 dimensional fermion system in the presence of
magnetic field.
There we expect Landau levels to contribute to the spectral density (Fig. 2).
 
In Section III we review the holographic description of the D3/Dp brane systems.
The subject of our interest is the fermion matter, which
 is formed (at finite chemical potential for the fermion number) in the low energy 
 theory living on intersection of the $N_c$ $D3$ branes 
 and $N_f$ Dp branes.
 We consider the case of $N_c\gg N_f\sim 1$ and strong 't Hooft coupling $\lambda$,
 where the holographic description is applicable.

In Section IV we focus on the zero sound mode  and show that
it develops a gap in the presence of magnetic field.
 In the case of vanishing magnetic field,  $B=0$, we observe a zero sound mode whose speed is the same as that of the first sound.
 As long as the value of the magnetic field $B$ is small compared to $\w^2,\q^2$ (in appropriate units), the sound mode peak
 in the spectral function is not significantly affected.
 On the other hand,  the presence of the nonvanishing magnetic field leads to a gap in the dispersion
 relation for zero sound. 
 (The effective action proposed by Nickel and Son \cite{Nickel:2010pr}  in the presence of the
 magnetic field gives vanishing sound velocity).
 In the regime of small magnetic field we derive the scaling behavior of the gap in the spectrum $\w_c$
as a function of magnetic field.
The result,  $\w_c\sim B$ is consistent with fermions acquiring an effective mass.

In Section V we investigate the current-current correlator at non-vanishing  frequency $\w$ and momentum $\q$.
We observe that in the longitudinal channel, the only nontrivial structure both
in the real and in imaginary parts of the correlators is provided by the zero sound.
There is no nontrivial structure in the transverse correlators when $B=0$.
We discuss our results  in Section VI.

In Appendix we consider higher derivative corrections and show that when they are added to the DBI the correlators are
not significantly modified.

\section{Fermi liquid and the random phase approximation}

In this section we  review the application of
the random phase approximation (RPA) for the computation of
the density-density response function $\langle J_0(\w,\q)J_0(\w,-\q)\rangle$
in Landau Fermi liquid theory. We  consider the 2+1 dimensional
theory for both cases of vanishing and non-vanishing magnetic field.

Due to the interaction of quasiparticles, the variation of
quasiparticle energy due to small perturbation of the distribution function,
is given by (see, e.g, \cite{Landau})
\beq
\delta\varepsilon ({\q})=\int d{\q}'f({\q},{\q}')\delta n({\q}')
\eeq
Because the small changes of quasiparticle density
occur in the vicinity of a Fermi surface, one considers the function
$f({\q},{\q}')$ to be dependent on the momenta on the Fermi
surface, and therefore it boils down to a function of the angle between
${\q}$ and ${\q}'$:
\beq
\frac{m^*}{\pi }f(\theta)=2F(\theta)\,.
\eeq
where, as usual, the effective mass at the Fermi surface is defined via
\beq
\label{meff}
    m^*=\frac{\q_F}{\upsilon_F}, \qquad \upsilon_F=\frac{\p \epsilon(\q)}{\p \q} |_{\q=\q_F}  
 \eeq
Landau parameters $F_l$ are the coefficients of the expansion of $F(\theta)$ in Legendre polynomials:
\beq
F(\theta)=\sum _l(2l+1)F_lP_l(\cos\theta)
\eeq
The Fermi liquid has a collective excitation at vanishing temperature called zero sound.
In the case of $F_l=0,\; l>0$, the speed of zero sound $u_0$ can be determined from
\beq
\frac{s}{2}\log\frac{s+1}{s-1}-1=\frac{1}{F_0},\, \qquad s=\frac{u_0}{\upsilon_F}    \label{Land-zero-sound-1}
\eeq
which, in the limit $F_0\gg1$ gives $s\sim \sqrt{F_0}$.

To compute the dynamical collective responses of a Fermi liquid,
one evaluates the time dependent mean field (random phase approximation)
obtained by summing up the quasiparticle ``bubble" diagrams.
Assuming for simplicity only the presence of a contact interaction,
with effective coupling constant $V\simeq F_0$, the $n${\it th} diagram is
equal to $V^{n-1}(\chi _0(\q,\w))^n$.
The susceptibility in the RPA is then given by the sum of a geometric progression:
\beq
\chi (\q,\w)=\frac{\chi _0(\q,\w)}{1-V\chi _0(\q,\w)}\,,
\eeq
Express $\chi =\chi ^{(r)}+i\chi ^{(i)}$, hence
\beq
\chi ^{(i)}(\q,\w)=\frac{\chi _0^{(i)}(\q,\w)}{(1-V\chi _0^{(r)}(\q,\w))^2+(\chi _0^{(i)}(\q,\w))^2}\,.
\eeq

\begin{figure}[htp]
\includegraphics[width=80mm]{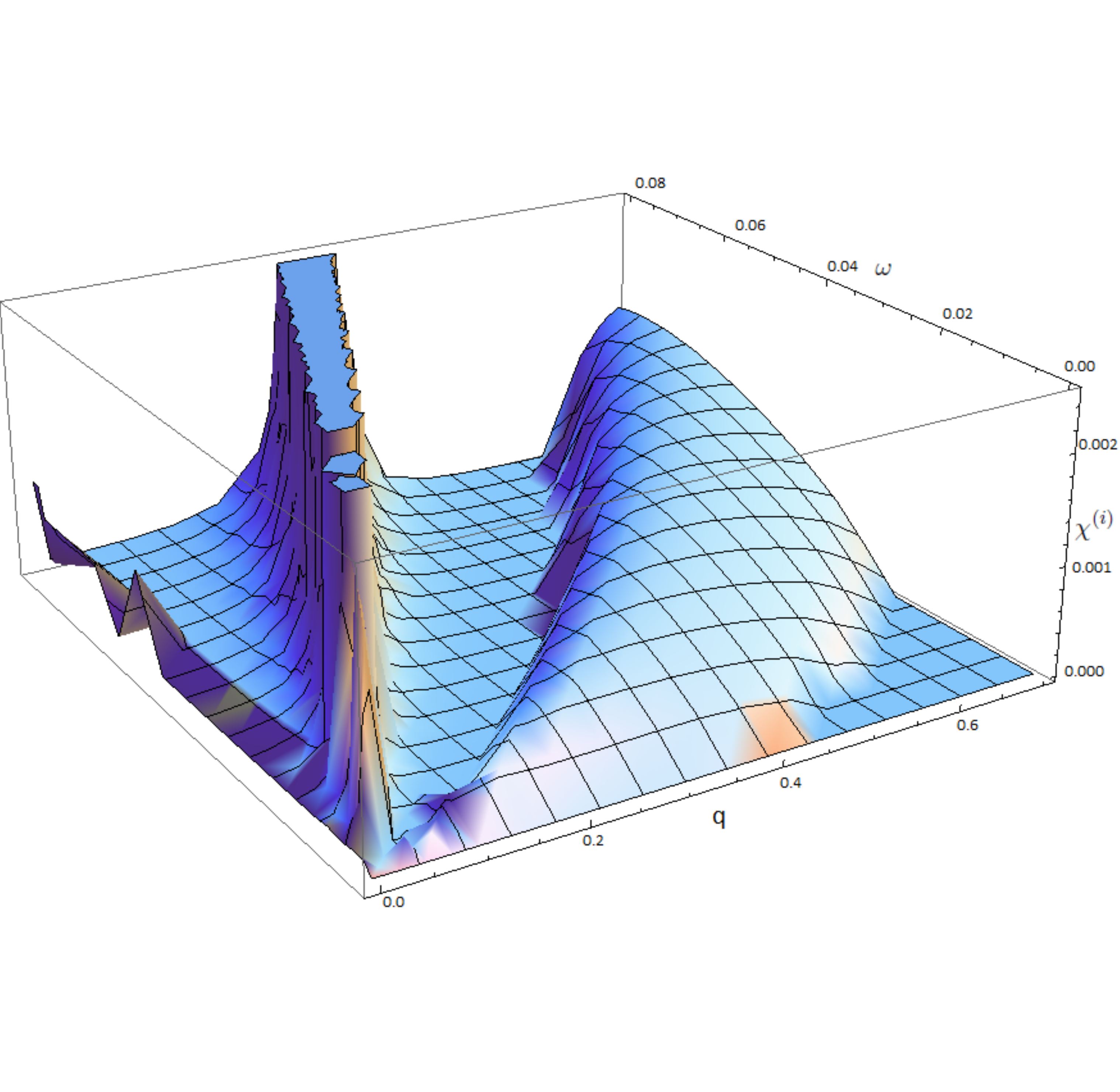}
\includegraphics[width=80mm]{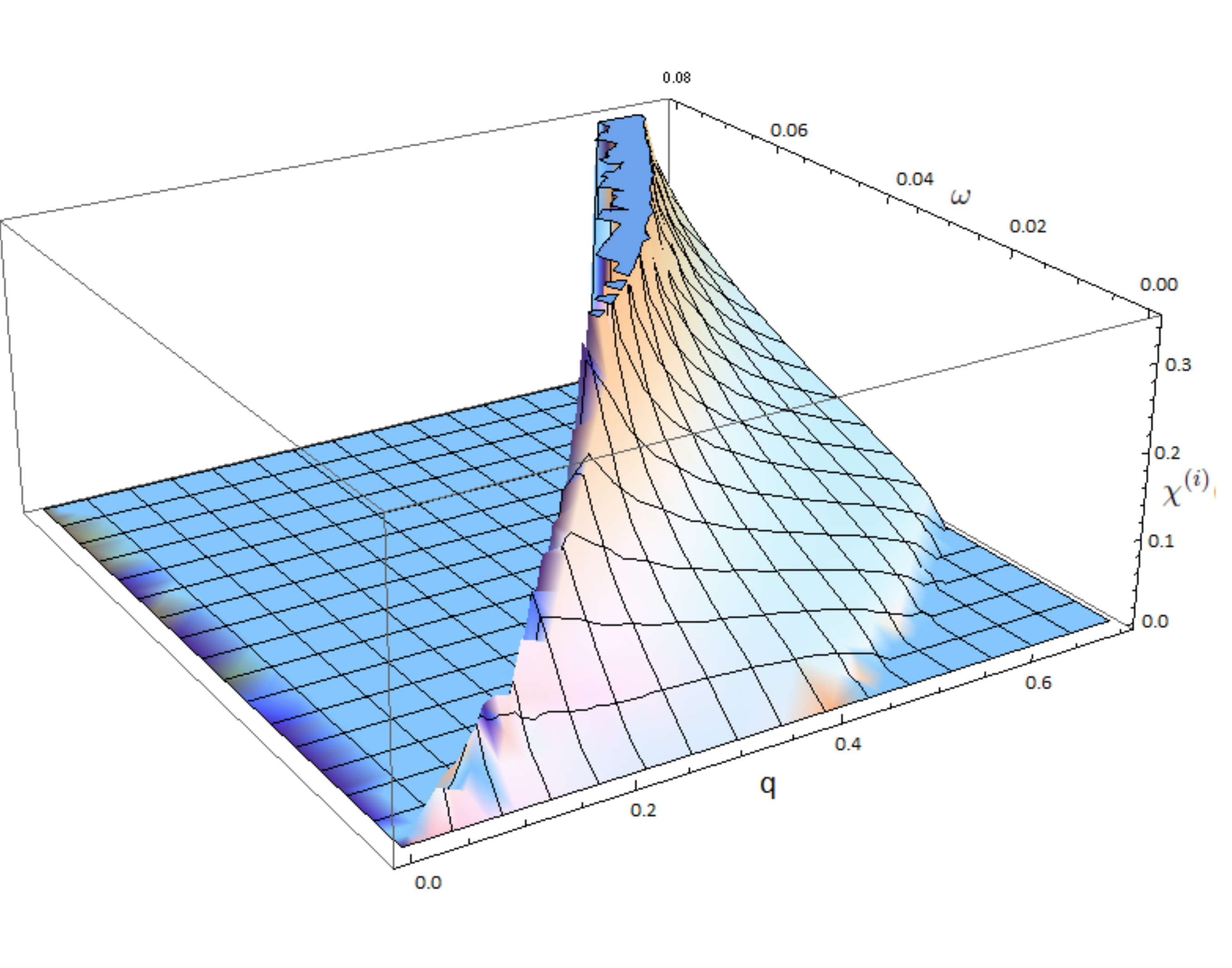}
\caption{Spectral density $\chi ^{(i)}(\q,\w)$ at strong coupling ($V=50$, left graph) and weak coupling
($V=3$, right graph) in the random phase approximation, at vanishing magnetic field. Fermi momentum is put to $q_F=0.2$. Note that at strong coupling
zero sound is well separated from the particle-hole continuum, while at weak coupling zero
sound merges with the left edge of the particle-hole continuum. At small frequencies particle-hole continuum
sharply ends at $\q=2\q_F$.}\label{b0rpa}
\end{figure}

Then we study density of excitations by plotting $\chi ^{(i)}(\q,\w)$.
The result for vanishing magnetic field is presented in Fig. \ref{b0rpa}, where
we plot the susceptibility (for $q_F=0.2$) at strong and weak coupling $V$.
In the case of strong coupling there is a finite gap, separating
the zero sound collective mode, and the band of the particle-hole excitations.
For given small frequency $\w$
the width of the gap is given by $\delta \q\simeq\frac{\w}{u_0}(s-1)$.
Note the non-analytic step behavior at $\q=2\q_F$, originating from the free response function $\chi _0^{(i)}(\q,\w)$.
In the case of weak coupling the zero sound mode merges with the left edge of particle-hole band.

The location of zero sound pole is determined as a solution to equations $\chi _0^{(i)}(\q,\w)=0$,  $\chi _0^{(r)}(\q,\w)=1/V$.
The real part $\chi _0^{(r)}(\q,\w)$ of Lindhard function for 2D Fermi gas is given by (see, e.g., \cite{giuliani}):
\beq
\chi _0^{(r)}(\q,\w)=-\left(1+\frac{\q_F}{\q}\left[\text{sign}(\nu _-)\theta (|\nu _-|-1)
\sqrt{\nu _-^2-1}-\text{sign}(\nu _+)\theta (|\nu _+|-1)\sqrt{\nu _+^2-1}\right]\right)\,,
\eeq
where $\nu _\pm=\frac{\w \pm\varepsilon _{\q}}{\q\upsilon _F}$. For large $\frac{\w}{\q \upsilon _F}=s\gg 1$ one may expand
\beq
\chi _0^{(r)}(\q,\w)\simeq\frac{\q^2\upsilon _F^2}{2\w^2}\,.\label{chi0b0-1}
\eeq
Therefore, for the speed of zero sound one obtains
$s=\sqrt{V/2}$, exactly as it follows at large $F_0$ from the equation \rref{Land-zero-sound-1}.

Suppose now that besides $F_0$ there is also non-vanishing ``mass" Landau parameter $F_1$.
In the relativistic case, the value of $m^*$ is related to the value of the chemical potential \cite{Baym:1975va},
\beq
\label{meffmu}
m^* = \mu \left( 1+\frac{F_1}{3} \right)
\eeq
The speed of zero sound $u_0$ then satisfies equation
\beq
\frac{s}{2}\log\frac{s+1}{s-1}-1=\frac{1+F_1/3}{F_0+F_0F_1/3+F_1s^2},\, \qquad s=\frac{u_0}{\upsilon_F}   \label{Land-zero-sound}
\eeq\\

For free fermions in a magnetic field $B$, the Lindhard function is equal to (see, e.g., \cite{giuliani}) 
\beq
\chi _0(\q,\w)=\frac{1}{2\pi\ell^2}\sum _{n,n'}\frac{f(\varepsilon _n)-f(\varepsilon _{n'})}{\w +(n-n')\w _c+i\eta}|F_{n',n}(\q)|^2\,,
\label{chi0bn0}
\eeq
where
\beq
F_{n',n}(\q)=\sqrt{\frac{n!}{n'!}}\left(\frac{(\q_y-i\q_x)\ell }{\sqrt{2}}\right)^{n'-n}e^{-\q^2\ell ^2/4}L_{n}^{n'-n}\left(\frac{\q^2\ell^2}
{2}\right)\,,\label{chi0bn0-F}
\eeq
for $n'\geq n$.
Here we have introduced the cyclotron frequency $\w _c=B/m^\star$ and the magnetic length $\ell=\frac{1}{\sqrt{B}}$.
The functions $L_{n}^{n'-n}$
are Laguerre polynomials, and $f(\varepsilon _n)$ is an occupation number for the $n${\it th} Landau level.

We would like to compute the effect of the magnetic field on the density-density response function
of the interacting fermions.
Let us write the quasiparticle interaction Hamiltonian
\beq
H_{int}=\sum _{{\bf q}}V_{{\bf q}}n_{{\bf q}}n_{-{\bf q}}\label{rpaHint}
\eeq
in the basis of Landau levels wavefunctions. The corresponding matrix elements of the density fluctuation operator
$n_{{\bf q}}=\sum _{{\bf k}}c_{{\bf k}}c_{{\bf k}+\q}^\dagger$ are given by
\beq
\langle n'{\bf k}_y'|n_\q |n{\bf k}_y\rangle =
\exp\left(-i\frac{\q_x ({\bf k}_y+{\bf k}^\prime_y)\ell^2}{2}\right)F_{n'n}(\q)\delta _{{\bf k}_y-{\bf k}^\prime_y,
\q_y}\,.
\eeq
The density fluctuation operator in the basis of Landau level wavefunctions is then given by
\beq
n_\q=\sum _{n,{\bf k}_y,\;n',{\bf k}^\prime_y} \langle n'{\bf k}_y'|n_\q |n{\bf k}_y\rangle c_{n{\bf k}_y}c_{n'{\bf k}^\prime _y}^\dagger
\label{nqLwg}
\eeq
Note that
\beq
\left(\langle n{\bf k}_y|n_\q |n'{\bf k}_y'\rangle\right)^\star =\langle n'{\bf k}_y'|n_{-\q} |n{\bf k}_y\rangle
\eeq
implies $(n_{\q})^\dagger =n_{-\q}$.
Substituting \rref{nqLwg}
into the interaction Hamiltonian \rref{rpaHint}, assuming again only a contact interaction of plane waves $V_\q\equiv V\simeq F_0$,
and considering all quasiparticles in the same Landau level $n$, one obtains
\beq
H_{int}=V\sum _{\q,{\bf k}_y,{\bf k}^\prime _y}c_{n{\bf k}_y}c^\dagger _{n{\bf k}_y-\q_y}c_{n{\bf k}^\prime_y}
c^\dagger _{n{\bf k}^\prime_y+\q_y}\exp\left(-i\ell ^2\q_x({\bf k}_y-{\bf k}_y^\prime -\q_y)-\frac{\q^2\ell^2}{2}\right)
[L_n^0(\q^2\ell^2/2)]^2\,.\label{Hintbn0}
\eeq
Let us choose the momentum to be in $y$-direction, then
\beq
H_{int}=\sum _{\q _y,{\bf k}_y,{\bf k}^\prime _y}V_{\q_y}c_{n{\bf k}_y}c^\dagger _{n{\bf k}_y-\q_y}c_{n{\bf k}^\prime_y}
c^\dagger _{n{\bf k}^\prime_y+\q_y}\,,
\eeq
where $V_{\q_y}=[L_n^0(\q^2\ell^2/2)]^2\exp\left(-\frac{\q^2\ell^2}{2}\right)V$.

\begin{figure}[htp]
\includegraphics[width=80mm]{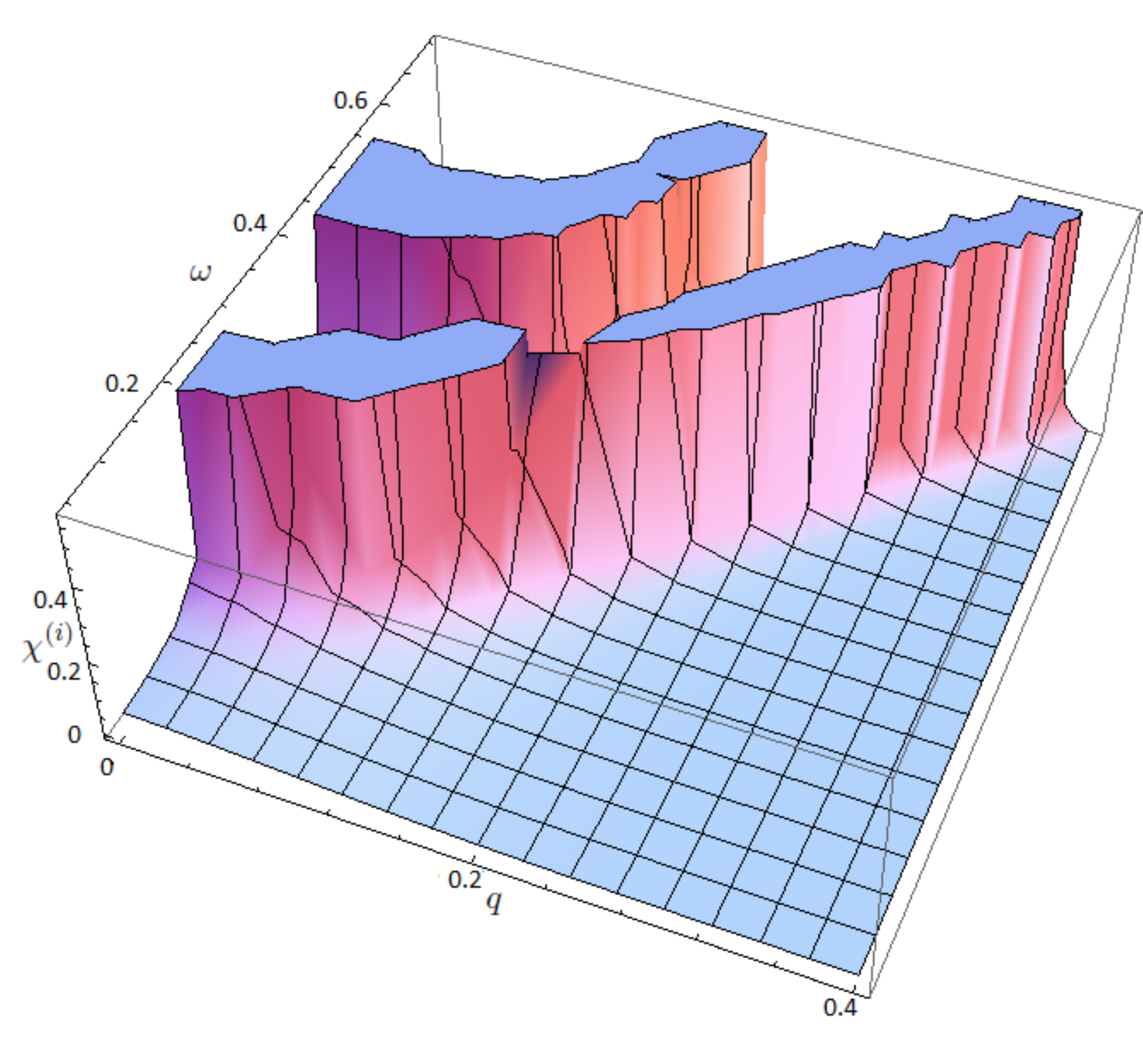}
\caption{Spectral density in the random phase approximation of the $2+1$ dimensional Fermi liquid
in the plane of the magnetic field, with cyclotron frequency $\omega_c=0.25$. First two of infinitely many collective excitation
branches are shown. Each branch starts at $(q=0,\,\omega=M\omega_c)$, where $M$
is an integer.}\label{rpaLandau}
\end{figure}

We can explicitly demonstrate that the zero sound mode is gapped in the magnetic field,
with the gap being equal to $\w _c$,
in agreement with the  Kohn's theorem \cite{Kohn:1961zz}.
For this aim we are to solve equation
$\chi _0^{(r)}(\q,\w)=1/V_{\q}$ again. From \rref{chi0bn0}, \rref{chi0bn0-F} one may obtain the following expression for $\chi _0^{(r)}$:
\beq
\chi _0^{(r)}(\q,\w)=\frac{e^{-\q^2\ell ^2/2}}{2\pi\hbar\ell ^2}\sum _{k=1}^\infty{\sum _j}^\prime\frac{j!}{(j+k)!}\left(\frac{\q^2\ell^2}{2}\right)^k
\left[L_j^k\left(\frac{\q^2\ell^2}{2}\right)\right]^2\frac{2k\w_c}{\w^2-(k\w _c)^2}\,,
\eeq
where the prime denotes summation in the range $\text{max}(0,\nu -k)\leq j\leq\nu$, and $\nu$ is the number of occupied Landau levels.
Following \cite{giuliani}, we consider this equation for small $\q$ and $\w\simeq \w _c$.
Then the main contribution in the sum over $k$ comes from the term with $k=1$, and we obtain equation: 
\beq
const\frac{\q^{2}}{\w^2-\w _c^2}\simeq\frac{1}{V}\,,
\eeq
and therefore the zero sound dispersion relation is given by
\beq
\w =\sqrt{\w _c^2+c \q^2}\,,
\eeq
where $c\sim V\w _c$ is a constant.
Similarly, for any integer $M$, there is a mode with dispersion relation
\beq
\w =\sqrt{(M\w _c)^2+c' \q^{2M}}\,.
\eeq
We plot RPA computations of two-point function, for $\omega _c=0.25$, restricting to the first two first branches,
in Fig. \ref{rpaLandau}.

\section{$Dp$ brane in $AdS_5\times S^5$ background}

We study strongly interacting massless fermions at zero temperature and finite density.
A good field theoretical model of such a
system is ${\cal N}=4$ SYM theory with gauge group $SU(N_c)$, coupled to matter in the fundamental representation.
A convenient way to study strongly coupled theories is provided by holography where
one considers a dual gravitational theory, taking the limit of large 't~Hooft coupling $\lambda=g_{YM}^2N_c$,
and the limit of large $N_c$.
The dual gravitational background is created by $N_c\gg 1$ $D3$ branes, and has an $AdS_5\times S^5$ geometry.
The coupling to fundamental matter is realized by considering an embedding
of a probe $Dp$ brane in the $AdS_5\times S^5$ background \cite{Karch:2002sh}.
We will consider $D3/Dp$ configurations
with $d=2+1$ dimensional intersections.

Let us now provide a more detailed description of the bulk gravitational theory set-up.
Consider $AdS_5\times S^5$ geometry, with the metric
\beq
ds^2=L^2\left(r^2(-dt^2+dx_\alpha dx^\alpha)+\frac{dr^2}{r^2}+d\Omega _5^2\right)\,.\label{AdS5-geom}
\eeq
Here $L$ is the radius of $S^5$ and scale of curvature of $AdS_5$.
We will study the probe $Dp$ brane, embedded in the geometry described by \rref{AdS5-geom}.
We represent the metric on $S^5$ as
\beq
d\Omega _5^2=d\Omega _n^2+\sin ^2\tilde\theta d\Omega _{5-n}^2=d\theta ^2+\sin ^2\theta d\Omega _{n-1}^2+\cos ^2\theta d\Omega _{5-n}^2\,,\nn
\eeq
where $n=p+1-d$.
Then we define coordinates $\rho\,,f$ via the relation
\beq
\rho =r\sin\theta\,,\quad\quad f=r\cos\theta\,,\quad r^2=\rho ^2+f^2\,,
\eeq
and write
\beq
d\theta ^2=\frac{(f-\rho\,\p_\rho f)^2}{r^4}d\rho ^2\,,\quad dr^2=\frac{(\rho +f\,\p_\rho f)^2}{r^2}d\rho ^2\,,
\eeq
which gives the following induced $Dp$ brane world-volume metric
\beq
ds_{Dp}^2=L^2\left(r^2(-dt^2+dx_idx^i)+\frac{1}{r^2}\left(1+(\p_\rho f)^2\right)d\rho ^2
+\frac{\rho^2}{r^2}d\Omega _{n-1}^2\right)\,.
\eeq
The coordinate $f(\rho)$ defines an embedding of the
$Dp$ brane in the AdS background \rref{AdS5-geom}. In the case of the trivial embedding $f(\rho)\equiv 0$,
which is what we are going to deal with in this paper, $Dp$ brane
crosses the Poincar\' e horizon of the AdS space. In the case of $d=3$ $p=7$ such a configuration
becomes stable only for sufficiently large values
of chemical potential $\bar\mu _{ch}$ in the dual field theory \cite{Kutasov:2011fr}.
(See also \cite{Evans:2011mu} for the phase structure of the similar model in the 
presence of the magnetic field.)
Note that holographically computed correlators do not depend on the dimensionality of
the probe brane; in particular our results apply in the case of stable supersymmetric D3/D5
defect theory. 
    
Subsequently we add a gauge field $A_\mu$ on the world-volume of the probe $D7$ brane.
In general we are interested in non-vanishing magnetic field $B$.
So we consider the following components of the field strength:
\beq
F_{12}=B\,,\quad\quad F_{0\rho}=-\p_\rho A_0(\rho)\,.
\eeq
Consequently the DBI action for the $Dp$ brane is given by
\footnote{We adopt the convention $2\pi\alpha '=1$. For our purposes we are ignoring the total numerical coefficient,
which leaves us with an overall normalization of the action proportional to $\frac{1}{g_s}\sim
\frac{N_c}{\lambda}\sim\frac{N_c}{L^4}$.}
\beq
S_{DBI}\simeq\frac{N_c}{L^4}\int d^{p+1}x\sqrt{-\det (G+F)}=\int d\Omega _{n-1}\int d^dx\, S\,,
\eeq
where we have denoted
\beq
S\simeq N_cL^{p-5}\int d\rho\rho ^{d-3}\sqrt{(L^4\rho ^4+B^2)(1-(\p_\rho A_0)^2L^{-4})}\,.\label{ActDBI}
\eeq
Now rescale gauge field on the world-volume as 
\beq
\bar A_\mu=\frac{A_\mu}{L^2}\,,\label{A-resc}
\eeq
which yields the DBI action in the form,
\beq
S\simeq N_cL^{p-3}\int d\rho\rho ^{d-3}\sqrt{(\rho ^4+\bar B^2)(1-(\p_\rho\bar A_0)^2)}\,,\label{Act-1}
\eeq
where $\bar B=B/L^2$.

In the case of a non-vanishing magnetic field there is also a Chern-Simons term in the total action for the $Dp$ brane. It can be shown that this term
vanishes in the case of $f\equiv 0$ embedding.

The boundary value of $\bar A_0$ is equal to the chemical potential of the dual field theory:
$\bar A_0(\rho=\infty)=\bar\mu _{ch}$.
Due to $f(\rho =0)=0$ and the initial condition $\bar A_0(r=0)=0$
(imposed to ensure that chemical potential vanishes when the charge density is zero)
we obtain $\bar A_0(\rho =0)=0$, and therefore the chemical potential may be expressed as
\beq
\bar\mu _{ch}=\int _0^\infty d\rho\,\p_\rho\bar A_0\,.\label{BHmu-ch}
\eeq
Introducing a constant of integration $\hat d$, the solution
of the equation of motion for $\p_\rho\bar A_0$ field strength becomes,
\beq
\p_\rho\bar A_0=\frac{\hat d^2}{\sqrt{\hat d^4+\rho ^4+\bar B^2}}\,.\label{A0backgr}
\eeq
Using this expression and eq. \rref{BHmu-ch}, we obtain the value of the chemical potential
\beq
\bar\mu _{ch}=\int _0^\infty d\rho\,\p_\rho\bar A_0=\frac{4\Gamma (5/4)^2}{\sqrt{\pi}}\frac{\hat d^2}{ (\hat d^4+\bar B^2)^{1/4}}\,.\label{much-d}
\eeq

\section{Holographic zero sound}

In this and the next sections we  study $D3/Dp$ system with $d=2+1$ dimensional intersection,
described by trivial $f(\rho)\equiv 0$ embedding of the probe $Dp$ brane in the $AdS_5\times S^5$ background.
We consider the gauge field on the $Dp$ brane world-volume, solve its classical equations of motion
and use AdS/CFT to find the two-point functions of the $U(1)$ current in the dual field theory.
In this section we
show the existence of holographic zero sound
in the $D3/Dp$ configuration, to observe that it develops a gap as the magnetic field
is turned on.
In the next section we will study the current-current correlation function numerically.

\subsection{Zero sound in the $D3/Dp$ system with $d=2+1$ dimensional intersection}

Equation \rref{A0backgr} is the expression for the background field strength $\p_\rho\bar A_0$.
Let us turn on small fluctuations $\bar a_0,\,\bar a_1,\,\bar a_2$,
dependent on coordinates $x^0,\,x^2,\,\rho$.
In addition let us fix the gauge $\bar a_\rho =0$.
The longitudinal response is described holographically by the
$\bar a_0$ and $\bar a_2$ components of the gauge field, and the transverse response is described by the $\bar a_1$ component.
The DBI action, expanded up to the second order in fluctuations, then takes the form
\footnote{We thank J. Shock for comments on this action.}
\begin{align}
S&=\int d\rho\left(\sqrt{\frac{\rho^4+\bar B^2}{1-(\p_\rho\bar A_0)^2}}\left(-\frac{(\p_\rho\bar a_0)^2}{1-(\p_\rho\bar A_0)^2}
+\frac{\rho^4(\p_\rho\bar a_2)^2-(\p_0\bar a_2-\p_2\bar a_0)^2}{\rho^4+\bar B^2}\right)+\right.\notag\\
 &+\sqrt{\frac{1-(\p_\rho\bar A_0)^2}{\rho^4+\bar B^2}}\left(\frac{\rho^4(\p_2\bar a_1)^2}{\rho^4+\bar B^2}
+\frac{\rho ^4(\p_\rho\bar  a_1)^2-(\p_0\bar a_1)^2}{1-(\p_\rho\bar  A_0)^2}\right)+\label{act-quad}\\
&+\left.\frac{2\bar B\p_\rho\bar  A_0}{\sqrt{(\rho^4+\bar B^2)(1-(\p_\rho\bar  A_0)^2)}}(\p_2\bar a_1\p_\rho\bar  a_0-
\p_0\bar a_1\p_\rho\bar  a_2+(\p_0\bar a_2-\p_2\bar a_0)\p_\rho\bar  a_1)\right)\notag
\end{align}
Note that the last line in \rref{act-quad} describes a coupling of the transverse and longitudinal
gauge potential components. Bellow we will consider Fourier transform of the gauge field
\beq
\bar a_\mu (\rho ,x^0,x^2)=\int\frac{d{\w} d{\q}}{(2\pi)^2}e^{-i{\w} x_0+i{\q}x_2}\tilde a_\mu (\rho,{\w}, {\q})
\label{fl-four}
\eeq
Now we substitute eq. \rref{A0backgr} into the action \rref{act-quad},
define $b^2=\bar B^2/\hat d^4$, and introduce a new
variable $z=\frac{\hat d}{\rho}$, so that  $z=0$ is a boundary and $z=\infty$ is a Poincar\'e horizon of $AdS_5$.
In addition, we make the quantities $\w,\,\q$ dimensionless, by measuring these in units of $\hat d$: $\w=\omega\hat d$, $\q=q\hat d$.
We also denote for shortness of notation
\beq
\zeta =1+(1+b^2)z^4
\eeq
Then the action \rref{act-quad} becomes written as
\begin{align}
S&=\int\frac{ dz}{1+b^2z^4}\left(-\zeta^{3/2}a_0^{\prime 2}+\zeta^{1/2}a_2^{\prime 2}-\zeta^{1/2}(\p_0a_2-\p_2a_0)^2+\zeta^{-1/2}(\p_2a_1)^2
-\zeta^{1/2}(\p_0a_1) ^2+\zeta^{1/2}a_1^{\prime 2}-\right.\notag\\
&\left.-2bz^4(\p_2a_1a_0'-\p_0a_1a_2'+(\p_0a_2-\p_2a_0)a_1')\right)\label{act-quad-111}\,,
\end{align}
where we have omitted bars for simplicity of notation, and prime denotes differentiation w.r.t. $z$.
In momentum representation
\begin{align}
S&=\int\frac{ dz}{1+b^2z^4}\left(-\zeta^{3/2}a_0^{\prime }(\omega,\,q)a_0^{\prime }(-\omega,\,-q)
+\zeta^{1/2}a_2^{\prime }(\omega,\,q)a_2^{\prime }(-\omega,\,-q)+\zeta^{1/2}E(\omega,\,q)E(-\omega,\,-q)+\right.\notag\\
&\left.+\zeta^{-1/2}q^2a_1(\omega,\,q)a_1(-\omega,\,-q)
-\zeta^{1/2}\omega ^2a_1(\omega,\,q)a_1(-\omega,\,-q)+\zeta^{1/2}a_1^{\prime }(\omega,\,q)a_1^{\prime }(-\omega,\,-q)+\right.\label{qactmr}\\
&\left. +2ibz^4(qa_1(-\omega,\,-q)a_0'(\omega,\,q)+\omega a_1(-\omega,\,-q)a_2'(\omega,\,q)+E(\omega,\,q)a_1'(-\omega,\,-q))\right)\,,\notag
\end{align}
where we have omitted tildes for simplicity of notation and introduced the gauge-invariant electric field strength \cite{Kovtun:2005ev},
\beq
E(\omega,\,q)=\omega a_2(\omega,\,q)+q a_0(\omega,\,q)\,.\label{E-def}
\eeq
In addition we have Gauss's law
\footnote{This is an equation of motion for $a_z$. To derive it replace
\beq
a_2'\rightarrow a_2'-\partial _2a_z\,,\quad a_0'\rightarrow a_0'-\partial _0a_z\nn
\eeq
in the Lagrangian \rref{act-quad-111} and leave only terms linear in derivatives of $a_z$, because only these will survive when we consider the
equation of
motion for $a_z$ in the $a_z =0$ gauge. Then use the Fourier transform \rref{fl-four}.
}
\beq
\omega \zeta ^{3/2}a_0'(\omega,\,q)+q\zeta ^{1/2}a_2'(\omega,\,q)=0\label{gauss-f}
\eeq
Together with
\beq
E'(\omega,\,q)=\omega a_2'(\omega,\,q)+qa_0'(\omega,\,q)\,,
\eeq
eq. \rref{gauss-f} gives
\beq
a_0'(\omega,\,q)=\frac{q}{q^2-\zeta\omega ^2}E'\,,
\eeq
\beq
a_2'(\omega,\,q)=\frac{\omega\zeta}{\omega^2\zeta-q^2}E'\,.
\eeq
Plugging these expressions into the action \rref{qactmr}, we obtain
\beq
S=\int\frac{ dz}{1+b^2z^4}\left(\frac{q^2-\zeta\omega ^2}{\zeta^{1/2}}a_1^2
-\zeta^{3/2}\frac{E^{\prime 2}}{\zeta\omega^2-q^2}+\zeta^{1/2}E^2+\zeta^{1/2}a_1^{\prime 2}+
2ibz^4(Ea_1)'\right)\,.\label{act-2}
\eeq
Corresponding fluctuation equations are
\begin{align}
E''&+\frac{2}{z}\left(\frac{1}{1+((1+b^2)z^4)^{-1}}+2\left(\frac{1}{1+b^2z^4}-\frac{1-(q/\omega)^2(1+(1+b^2)z^4)^{-2}}{1-(q/\omega)^2(1+(1+b^2)z^4
)^{-1}}\right)\right)E'+ \notag \\
&+\left(\omega ^2-\frac{q^2}{1+(1+b^2)z^4}\right)E-\frac{4ibz^3(\omega^2(1+(1+b^2)z^4)-q^2)a_1}{
(1+b^2z^4)(1+(1+b^2)z^4)^{3/2}}=0\label{fleqgen-1}
\end{align}
\begin{equation}
a_1''+2z^3\left(\frac{1+b^2}{1+(1+b^2)z^4}-\frac{2b^2}{1+b^2z^4}\right)a_1'+
\left(\omega ^2-\frac{q^2}{1+(1+b^2)z^4}\right)a_1+\frac{4ibz^3E}{(1+b^2z^4)(1+(1+b^2)z^4)^{1/2}}=0\label{fleqa1bnon0}
\end{equation}

\subsubsection{Vanishing magnetic field}

In this subsection we set the magnetic field to zero.
Fluctuations of $E$ and $a_1$ fields then decouple, and we can consider separately
transverse and longitudinal responses,
\beq
E''+\frac{2}{z}\left(\frac{1}{1+z^{-4}}+2\left(1-\frac{1-(q/\omega)^2(1+z^4)^{-2}}{1-(q/\omega)^2(1+z^4)^{-1}}\right)
\right)E'+(\omega ^2-q^2(1+z^4)^{-1})E=0\,,\label{fleqb0}
\eeq
\beq
a_1''+\frac{2z^3}{1+z^4}a_1'+\left(\omega ^2-\frac{q^2}{1+z^4}\right)a_1=0\,.\label{a1fleq}
\eeq
Let us first study the longitudinal response.
In the near-horizon $z\gg 1$ region eq. \rref{fleqb0} becomes:
\beq
E''+\frac{2}{z}E'+\omega ^2E=0\,,\label{B0zInf}
\eeq
The general solution of \rref{B0zInf} is a linear combination of $e^{\pm i\omega z}/z$.
We choose the solution with the incoming near-horizon behavior, since it corresponds to retarded propagator in the dual field theory \cite{Son:2002sd}:
\beq
E=C\frac{e^{i\omega z}}{z}\,.\label{E-z-1}
\eeq
The constant $C$ is undetermined, because the fluctuation equation is linear.
When $\omega z\ll 1$, we obtain
\beq
E=C\left(\frac{1}{z}+i\omega\right)\,.\label{E-z-11}
\eeq

Condition \rref{E-z-1} together with the boundary condition $E(0)=0$ (imposed to get normalizable  solutions)
defines an eigenvalue problem for the fluctuation equation \rref{fleqb0}.
In the limit $\omega z, q z\ll 1$, \rref{fleqb0} reduces to,
\beq
E''+\frac{2}{z}\left(\frac{1}{1+z^{-4}}+2\left(1-\frac{1-(q/\omega)^2(1+z^4)^{-2}}{1-(q/\omega)^2(1+z^4)^{-1}}\right)\right)E'=0\,,\label{hwq-1}
\eeq
having as general solution,
\beq
E(z)=C_1+C_2(q^2-2\omega ^2)\sqrt{i}F\left(i\sinh^{-1}(\sqrt{i}z)|-1\right)-\frac{C_2q^2z}{\sqrt{1+z^4}}\,,\label{E-z-2}
\eeq
where $F(z)$ is an elliptic integral of the first kind.
In the limit $z\rightarrow\infty$ it has an expansion
\beq
\sqrt{i}F\left(i\sinh^{-1}(\sqrt{i}z)|-1\right)\rightarrow -K(1/2)+\frac{1}{z}+O\left(\frac{1}{z^5}\right)\,,\label{ellint-as}
\eeq
where $K(z)$ is the complete elliptic integral of the first kind. The solution \rref{E-z-2} becomes in this limit
\beq
E(z)=C_1-C_2K(1/2)(q^2-2\omega ^2)-\frac{2C_2}{z}\omega ^2.\label{E-z-3}
\eeq
Now we compare \rref{E-z-11} and \rref{E-z-3}, and obtain as result
\beq
C_1=\left(i\omega -\frac{(q^2-2\omega ^2)K(1/2)}{2\omega ^2}\right)C\,,
\quad\quad C_2=-\frac{C}{2\omega ^2}
\eeq
Recalling the boundary condition $E(0)=0$, we deduce from \rref{E-z-2} that $C_1=0$, and consequently
\beq
\left(1+\frac{i\omega}{K(1/2)}\right)^{-1}=\frac{2\omega ^2}{q^2}\,,
\eeq
which in the limit of small $q,\,\omega$ is solved by the considering leading orders in momentum $q$,
\beq
\omega =\pm\frac{q}{\sqrt{2}}-\frac{iq^2}{4K(1/2)}\,.\label{zeosdis}
\eeq
This excitation has been identified before, and is called \cite{Karch:2008fa} holographic zero sound.
In the $d=2+1$ dimensional system this mode has been observed in \cite{Bergman:2011rf}.
Note that the speed of sound does not depend on dimensionality $p$ of a probe brane and for any value of $p$
is equal to the speed of the usual sound in the hydrodynamic regime  \cite{Karch:2009eb}.
In Section IV we will study current-current two-point functions,
and the peak in the spectral function, corresponding to zero sound mode, will also be observed in the numerics.

Now, let us consider the fluctuation equation \rref{a1fleq} for the transverse gauge field component, in the limit $\omega,\,q\ll 1$.
Then eq. \rref{a1fleq} becomes
\beq
a_1''+\frac{2z^3}{1+z^4}a_1'=0\,,
\eeq
with an exact solution being
\beq
a_1(z)=C_1+C_2\sqrt{i}F\left(i\sinh ^{-1}(\sqrt{i}z)|-1\right)\,.\label{a1smwq}
\eeq
In the near-horizon $z\rightarrow\infty$ limit it is expanded as
\beq
a_1(z)\simeq C_1-C_2K(1/2)+C_2/z\,.\label{a-z-3-1}
\eeq
Comparing it with the incoming-wave solution \rref{a1incwave}, one obtains
\beq
C_1=\left(i\omega +K(1/2)\right)C\,,\quad\quad C_2=C\,.
\eeq
Then, near-boundary $z\ll 1$ expansion of \rref{a1smwq} is given by
\beq
a_1(z)\simeq A+Bz\,,
\eeq
where
\beq
A=\left(i\omega +K(1/2)\right)C\,,\quad\quad B=-C\,.
\eeq
Therefore one may find the current two-point function $\langle J^1J^1\rangle =\frac{B}{A}$.
In particular, its imaginary part is given by
\beq
\text{Im}\langle J^1J^1\rangle \simeq\frac{\omega}{[K(1/2)]^2}\,.\label{Imj1j1a}
\eeq
We provide numerical results for the transverse fluctuations in Section IV.

\subsubsection{Non-vanishing magnetic field}

In this subsection we are going to study the case of small magnetic field, $b\ll 1$, which will allow
us to achieve some simplifications. Let us rewrite the action \rref{act-2} as
\beq
S=\int dz\left({\cal G}_EE^{\prime 2}+{\cal U}_EE^2+{\cal G}_a a_1^{\prime 2}+{\cal U}_a a_1^2+{\cal C}^{(1)}(Ea_1)'\right)\,,
\label{act-3}
\eeq
where we have denoted
\beq
{\cal G}_E=-\frac{(1+(1+b^2)z^4)^{1/2}}{(1+b^2z^4)\left(\omega ^2-\frac{q^2}{1+(1+b^2)z^4}\right)}\,,\quad
{\cal U}_E=\frac{(1+(1+b^2)z^4)^{1/2}}{1+b^2z^4}\,,
\eeq
\beq
{\cal G}_a=\frac{(1+(1+b^2)z^4)^{1/2}}{1+b^2z^4}\,,\quad
{\cal U}_a=-\frac{(1+(1+b^2)z^4)^{1/2}\left(\omega ^2-\frac{q^2}{1+(1+b^2)z^4}\right)}{1+b^2z^4}\,,
\eeq
\beq
{\cal C}^{(1)}=\frac{2ibz^4}{1+b^2z^4}\,.
\eeq
In the case of $b\ll 1$, we can approximate
\beq
{\cal G}_E=-\frac{(1+z^4)^{1/2}}{(1+b^2z^4)\left(\omega ^2-\frac{q^2}{1+z^4}\right)}\,,\quad
{\cal U}_E=\frac{(1+z^4)^{1/2}}{1+b^2z^4}\,,
\eeq
\beq
{\cal G}_a=\frac{(1+z^4)^{1/2}}{1+b^2z^4}\,,\quad
{\cal U}_a=-\frac{(1+z^4)^{1/2}\left(\omega ^2-\frac{q^2}{1+z^4}\right)}{1+b^2z^4}\,,
\eeq
\beq
{\cal C}^{(1)}=\frac{2ibz^4}{1+b^2z^4}\,.
\eeq

In the near-horizon limit, for $\omega >0$, integrating the ${\cal C}^{(1)}$ term by parts, we arrive at
\beq
S=\int\frac{ dz\,z^2}{1+b^2z^4}\left(-\frac{E^{\prime 2}}{\omega ^2}+E^2+a_1^{\prime 2}-(\omega ^2-b^2q^2)a_1^2
-8ibEa_1\frac{z}{1+b^2z^4}\right)\,.\label{nhl}
\eeq
Moreover, for $z\gg 1/\sqrt{b}$ and $z^3\gg 1/(b(\omega ^2-b^2q^2)^{1/2})$, we actually obtain decoupled system of equations
\beq
E''-\frac{2}{z}E'+\omega ^2E=0\quad\Rightarrow\quad E=\tilde C_1(1-i\omega z)e^{i\omega z}\,,\label{nhsol1}
\eeq
\beq
a_1''-\frac{2}{z}a_1'+(\omega ^2-b^2q^2)a_1=0\quad\Rightarrow\quad a_1=\tilde C_2(1-i\sqrt{\omega ^2-b^2q^2}z)e^{i\sqrt{\omega ^2-b^2q^2}z}\,.
\label{nhsol2}
\eeq
Now, assume that $\omega ^2\gg b^2q^2$, and perform the linear transformation in \rref{nhl}
\beq
E=i\omega (\chi_1\tilde E+\chi_2\tilde a_1)\,,\quad\quad a_1=\chi_1\tilde E-\chi_2\tilde a_1\,,
\eeq
with arbitrary constant coefficients $\chi _1,\,\chi_2$, 
which brings the action to the form
\beq
S=\int\frac{ 2dz\,z^2}{1+b^2z^4}\left(\chi_1^2\left(\tilde E^{\prime 2}-\omega ^2\left(1-\frac{4bz}{\omega(1+b^2z^4)}\right)\tilde E^2\right)
+\chi_2^2\left(\tilde a_1^{\prime 2}-\omega ^2\left(1+\frac{4bz}{\omega (1+b^2z^4)}\right)\tilde a_1^2\right)\right)
\eeq
Corresponding equations of motion are
\beq
\tilde E''+\frac{2}{z}\frac{1-b^2z^4}{1+b^2z^4}\tilde E'+\omega^2\left(1-\frac{4bz}{\omega(1+b^2z^4)}\right)\tilde E=0\,,
\eeq
\beq
\tilde a_1''+\frac{2}{z}\frac{1-b^2z^4}{1+b^2z^4}\tilde a_1'+\omega^2\left(1+\frac{4bz}{\omega(1+b^2z^4)}\right)\tilde a_1=0\,.
\eeq
The solutions are
\beq
\tilde E=\frac{e^{\pm i\omega z}}{z}+\frac{b}{\omega}(1\mp i \omega z)e^{\pm i\omega z}\,,
\eeq
\beq
\tilde a_1=\frac{e^{\pm i\omega z}}{z}-\frac{b}{\omega}(1\mp i \omega z)e^{\pm i\omega z}\,.
\eeq
We impose the incoming-wave behavior,
\beq
E=i\omega\left(\frac{\chi_1+\chi_2}{z}+(\chi_1-\chi_2)\frac{b}{\omega}(1-i\omega z)\right)e^{i\omega z}\,,\label{bnoenh}
\eeq
\beq
a_1=\left(\frac{\chi_1-\chi_2}{z}+(\chi_1+\chi_2)\frac{b}{\omega}(1-i\omega z)\right)e^{i\omega z}\,,\label{bnoanh}
\eeq
which leaves us with two constant of integration $\chi_1\pm\chi_2$.
\\

When $\omega\sim q\ll 1$, we can consider fluctuation equations  \rref{fleqb0}, \rref{a1fleq},
as for the case of vanishing magnetic field. Then we perform computations along the lines of the previous subsection,
using now near-horizon boundary conditions \rref{bnoenh} and \rref{bnoanh}.

First, we match \rref{bnoenh} in $\omega z\ll 1$ limit,
\beq
E=i(b(\chi_1-\chi_2)+i\omega ^2(\chi_1+\chi_2))+i\omega\frac{\chi_1+\chi_2}{z}
\eeq
with eq. \rref{E-z-3}. Requiring that $C_1=0$, we arrive at
\beq
q^2-2\omega ^2-\frac{2}{K(1/2)}\left(\omega b\frac{\chi_1-\chi_2}{\chi_1+\chi_2}+i\omega ^3\right)=0\,.\label{c1-e}
\eeq
Then, we match \rref{bnoanh} in $\omega z\ll 1$ limit,
\beq
a_1=i\omega (\chi_1-\chi_2)+(\chi_1+\chi_2)\frac{b}{\omega}+\frac{\chi_1-\chi_2}{z}
\eeq
with eq. \rref{a-z-3-1}. Again, imposing normalizability condition $C_1=0$, we obtain
\beq
\omega+\frac{1}{K(1/2)}\left(b\frac{\chi_1+\chi_2}{\chi_1-\chi_2}+i\omega ^2\right)=0\,.\label{c1-a1}
\eeq
Solving \rref{c1-e} together with \rref{c1-a1},
we get \footnote{Equivalently, we can obtain this result requiring that
\rref{c1-e} and \rref{c1-a1} have a non-trivial solution for $\chi_1\pm\chi_2$.}
\beq
q^2-2\omega ^2+\frac{2b^2}{[K(1/2)]^2}+\frac{i\omega}{K(1/2)}(q^2-4\omega ^2)=0\,.\label{gzsd}
\eeq
We see that in the presence of a magnetic field $b$ zero sound mode
develops a gap $\omega _c$ in the spectrum,
\beq
\omega _c=\frac{b}{K(1/2)}\,.
\eeq
\\

\subsection{Effective theory for the sound mode}

Zero sound may also be studied in the framework of Ref. \cite{Nickel:2010pr}.
First, one introduces a hypersurface $z=z_\Lambda$ in the bulk, integrating out degrees of freedom in the UV region $0\le z\le z_\Lambda$.
The UV physics is then effectively encoded in the action by,
\beq
S=\frac{1}{2}\int d^3x(f_0^2(\partial _0\phi -W_0+w_0)^2-f_2^2(\partial _2\phi -W_2+w_2)^2)\,,\label{NS-lagr}
\eeq
where $W_\mu =a_\mu(z=0)$, $w_\mu=a_\mu(z=z_\Lambda)$, and the ``Godstone boson" $\phi$ corresponds
to breaking of the $U(1)$ symmetry with a gauge field $W_\mu -w_\mu$.
The zero sound mode may be interpreted in such a framework as a mode coming from an
excitation of the field $\phi$, and therefore the speed of zero sound is given by the expression $v=f_2/f_0$.
Let us now compare the effective field theory action for the UV degrees of freedom with the bulk DBI action.
To render the relation between bulk and boundary  to be precise,
we specify the zero boundary condition $W_\mu=0$, putting the Goldstone boson $\phi$ to zero:
\beq
S=\frac{1}{2}\int d^3x(f_0^2 w_0^2-f_2^2 w_2^2)\label{bound-zL}
\eeq
Let us consider all fields to be only $z$-dependent, in which case transverse fluctuations decouple,
and we can put these to zero. Then
we can rewrite the bulk theory action \rref{act-quad-111} in a form
\beq
S\simeq\frac{1}{2}\int d^3x\frac{dz}{1+b^2z^4}(h^3(z)\tilde a_0^{\prime 2}-h(z)\tilde a_2^{\prime 2})\,,\label{fl-comp}
\eeq
where we have defined $h(z)=\sqrt{1+(1+b^2)z^4}$.
The solutions of the equations of motion on $\tilde a_0,\,\tilde a_2$, satisfying zero boundary condition at the AdS boundary,
while being defined on the hypersurface $z=z_\Lambda$, are now given by:
\beq
w_0=C_0\int _0^{z_\Lambda}\frac{dz(1+b^2z^4)}{h^3(z)}\,,\quad\quad w_2=C_2\int_0^{z_\Lambda}\frac{dz(1+b^2z^4)}{h(z)}\,.\label{fl-comp-sol}
\eeq
To match the bulk action and the boundary theory
\rref{bound-zL}, we evaluate the action \rref{fl-comp}
on the solution of the EOM, which leaves us with the boundary
terms at $z=z_\Lambda$ only
\beq
S\simeq\frac{1}{2}\int d^3x(C_0w_0-C_2w_2)\,,
\eeq
which in turn with the help of \rref{fl-comp-sol}, may be rewritten as \rref{bound-zL} with
\beq
f_0^{-2}=\int _0^{z_\Lambda}\frac{dz(1+b^2z^4)}{h^3(z)}\,,\quad\quad f_2^{-2}=\int _0^{z_\Lambda}\frac{dz(1+b^2z^4)}{h(z)}\,.
\eeq
Therefore the speed of zero sound is given by
\beq
u_0^2=\int _0^{z_\Lambda}\frac{dz(1+b^2z^4)}{h^3(z)}\left(\int _0^{z_\Lambda}\frac{dz(1+b^2z^4)}{h(z)}\right)^{-1}\,.\label{Son-v}
\eeq
When $b\ll 1$, one obtains
\beq
u_0^2\simeq\frac{1}{2+\frac{8\pi ^{1/2}}{3\Gamma[1/4]^2}b^2z_\Lambda ^3}\,,
\eeq
and therefore for $b^2z_\Lambda ^3\ll 1$ one recovers the value
of the speed of zero sound in vanishing b-field, $u_0=1/\sqrt{2}$, while for
$b^2z_\Lambda ^3\gg 1$ the speed of zero sound approaches zero.
In this regime the description of the low energy physics by the effective action  \rref{NS-lagr} presumably breaks down; it would be interesting
to write the low energy description that would account for the gap in the spectrum.


\subsection{Thermodynamic properties of trivial embeddings}

We will study the thermodynamics of the trivial $Dp$ brane embedding,
to obtain as a result the value of the speed of the usual first (hydrodynamic) sound. 
We consider here the $D3/Dp$ system with a $2+1$ dimensional intersection, and in the Appendix we will study the supersymmetric $D3/D7$
system with a $3+1$ dimensional intersection, in the presence of a non-vanishing magnetic field.

The total prefactor of the action is irrelevant for the computation of the speed of first sound.
The grand canonical potential is given by the equation
\beq
\Xi =-S=\int d\rho (\rho ^4+\bar B^2)(\rho ^4+\bar B^2+\hat d^4)^{-1/2}=a\frac{2\bar B^2-\hat d^4}{(\bar B^2+\hat d^4)^{1/4}}\,,\label{onsheactA}
\eeq
where $a=\Gamma (1/4)^2/(12\sqrt{\pi})$. Using \rref{much-d} one may calculate the charge density as,
\beq
\hat\rho =-\frac{\partial\Xi}{\partial\bar\mu _{ch}}\,,
\eeq
to find the energy density, being at zero temperature equal to the free energy,
\beq
\epsilon =\Xi +\bar\mu _{ch}\hat\rho =2a(\bar B^2+\hat d^4)^{3/4}\,.
\eeq
Consequently, the speed of sound is given by
\beq
u^2=\frac{\partial P}{\partial\epsilon}=-\frac{\partial \Xi}{\partial\epsilon}=\frac{1}{2}\frac{1+2b^2}{1+b^2}\,.
\eeq
Notice that this result is independent of $p$, which agrees with \cite{Karch:2009eb}.
Observe that when the magnetic field vanishes we retrieve the value $u^2=1/2$, which we observed before
in the dispersion relation \rref{zeosdis}.

Notice also that all the steps performed in the above may be combined into one expression
(use $\partial S/\partial\bar\mu _{ch}=\hat d^2$):
\beq
u^2=\frac{\partial S/\partial\bar\mu _{ch}}{\bar\mu _{ch}\partial ^2S/\partial\bar\mu _{ch} ^2}=\frac{1}{2}\frac{\partial\log\bar\mu _{ch}}
{\partial\log\hat d}\,.\label{therm-sound}
\eeq

\section{Holographic current-current correlators at finite frequency and momentum}

In the previous section we have shown that a propagating mode (zero sound)
develops a gap in the presence of the magnetic field.
In this section we  compute numerically the
two-point function of the $U(1)$ currents. First we set magnetic field to zero.
We  identify the holographic zero sound as a peak in the spectral function.
We start by computing the density-density correlator $\langle J^0J^0\rangle$
using the linearized DBI action.
We  then proceed to computing the transverse correlator $\langle J^1J^1\rangle$.
After that we proceed to the case of non-vanishing magnetic field and
show that the gap in the zero sound spectrum shows itself on the numeric graphs.

\subsection{Fluctuations of electric field strength $E$}

Consider the fluctuation equation \rref{fleqgen-1}, near the boundary $z=0$ for any value of magnetic field:
\beq
E''-( q^2-\omega ^2)E=0\,.\label{nsusyb}
\eeq
Its general solution is of the form,
\beq
E={\cal A}_EF_{I}+{\cal B}_EF_{II}\,,
\eeq
where we have denoted the two independent solutions as
\beq
F_I=1+\frac{ q^2-\omega ^2}{2} z^2+\frac{( q^2-\omega ^2)^2}{24} z^4+\cdots\,,\label{FI}
\eeq
\beq
F_{II}= z+\frac{ q^2-\omega ^2}{6} z^3+\cdots\,.\label{FII}
\eeq

The on-shell action is therefore given by
\beq
S_{on-shell}\simeq\lim_{\varepsilon\rightarrow 0}
\int d\omega dq{\cal A}_E(\omega,q){\cal A}_E(-\omega,-q)\frac{1}{q^2-\omega^2} \frac{{\cal B}_E(\omega,q)}{{\cal A}_E(\omega,q)}|_{z=\varepsilon}\,.
\eeq
Non-vanishing Green functions are 
\begin{align}
\langle J^0(\omega,q)J^0(-\omega,-q)\rangle&=\lim _{\varepsilon\rightarrow 0}
\frac{\delta ^2S_{on-shell}}{\delta a_0(z=\varepsilon,\omega,q)\delta a_0(z=\varepsilon,-\omega,-q)}=\notag\\
&=\lim _{\varepsilon\rightarrow 0}
\frac{\delta ^2S_{on-shell}}{\delta E(z=\varepsilon,\omega,q)\delta E(z=\varepsilon,-\omega,-q)}\frac{\delta E(\omega,q,z)}{\delta a_0(\omega,q,z)}
\frac{\delta E(-\omega,-q,z)}{\delta a_0(-\omega,-q,z)}=\label{j0j0a}\\
&=-\frac{q^2}{q^2-\omega ^2}\frac{ {\cal B}_E(\omega,q)}{ {\cal A}_E(\omega,q)}\,,\notag\\
\langle J^2(\omega,q)J^2(-\omega,-q)\rangle&=\lim _{\varepsilon\rightarrow 0}
\frac{\delta ^2S_{on-shell}}{\delta a_2(z=\varepsilon,\omega,q)\delta a_2(z=\varepsilon,-\omega,-q)}=\notag\\
&=\lim _{\varepsilon\rightarrow 0}
\frac{\delta ^2S_{on-shell}}{\delta E(z=\varepsilon,\omega,q)\delta E(z=\varepsilon,-\omega,-q)}\frac{\delta E(\omega,q,z)}{\delta a_2(\omega,q,z)}
\frac{\delta E(-\omega,-q,z)}{\delta a_2(-\omega,-q,z)}=\label{j2j2a}\\
&=-\frac{\omega^2}{q^2-\omega ^2}\frac{ {\cal B}_E(\omega,q)}{ {\cal A}_E(\omega,q)}\,,\notag\\
\langle J^0(\omega,q)J^2(-\omega,-q)\rangle&=\lim _{\varepsilon\rightarrow 0}
\frac{\delta ^2S_{on-shell}}{\delta a_2(z=\varepsilon,\omega,q)\delta a_0(z=\varepsilon,-\omega,-q)}=\notag\\
&=\lim _{\varepsilon\rightarrow 0}
\frac{\delta ^2S_{on-shell}}{\delta E(z=\varepsilon,\omega,q)\delta E(z=\varepsilon,-\omega,-q)}\frac{\delta E(\omega,q,z)}{\delta a_2(\omega,q,z)}
\frac{\delta E(-\omega,-q,z)}{\delta a_0(-\omega,-q,z)}=\label{j0j2a}\\
&=-\frac{\omega q}{q^2-\omega ^2}\frac{ {\cal B}_E(\omega,q)}{ {\cal A}_E(\omega,q)}\,.\notag
\end{align}

Note that these expression agree with the Ward identity for the $U(1)$ conserved current $J^\mu$,
\beq
\omega \langle J^0(\omega,q)J^0(-\omega,-q)\rangle -q\langle J^0(\omega,q)J^2(-\omega,-q)\rangle
=0\,.
\eeq

\begin{figure}[htp]
\includegraphics[width=80mm]{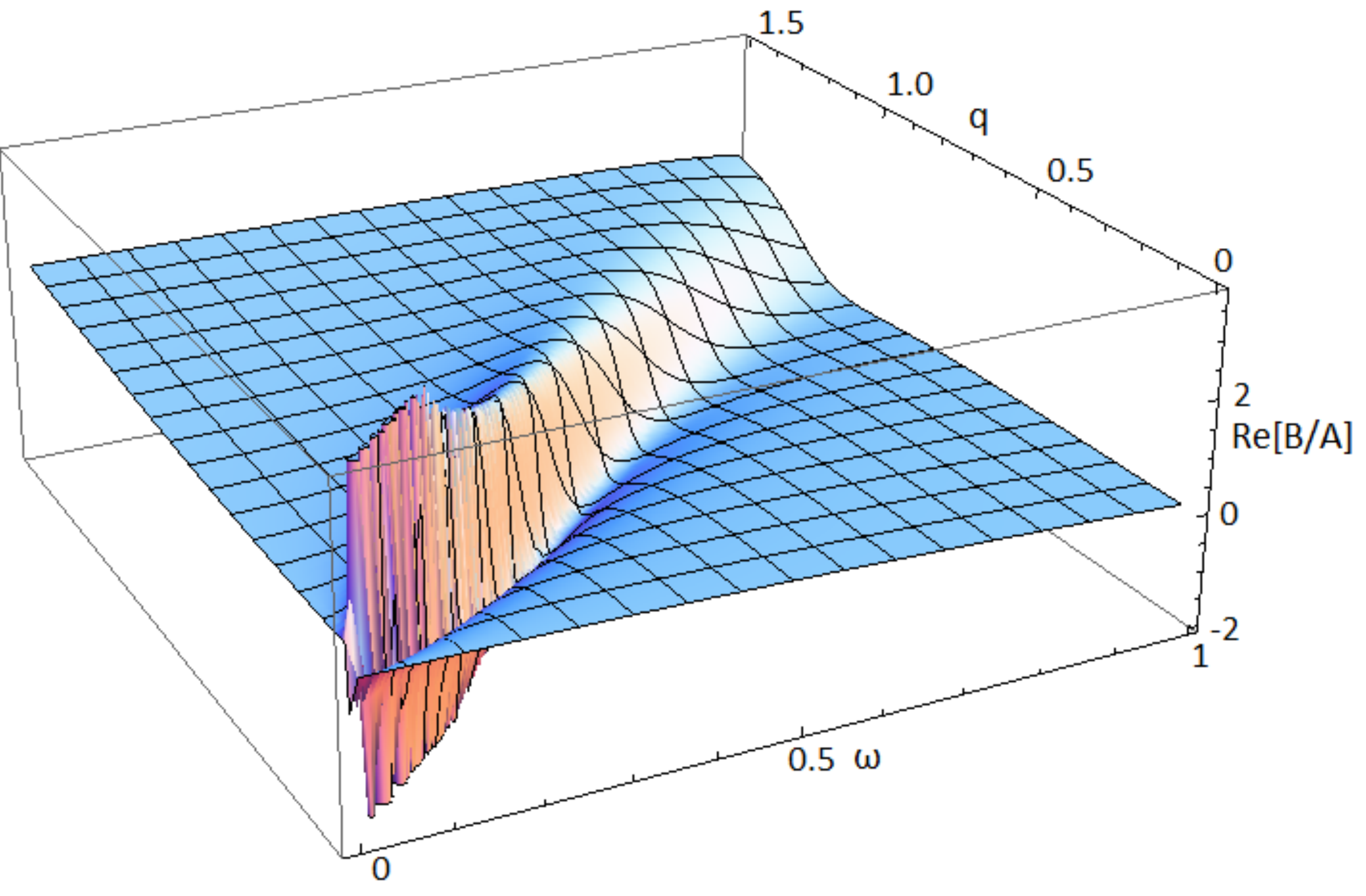}
\includegraphics[width=80mm]{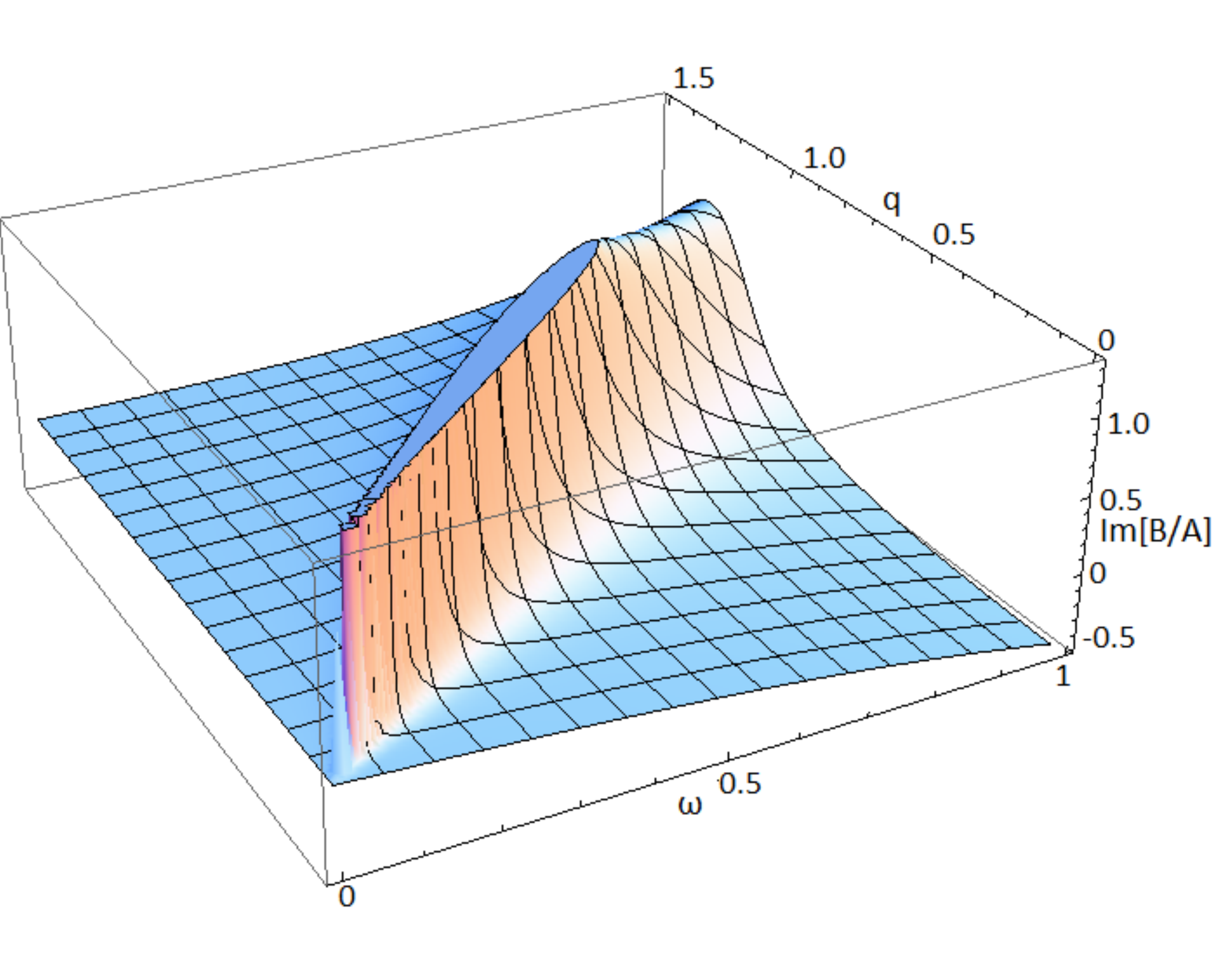}
\caption{Real and imaginary parts of the ${\cal B}_E/{\cal A}_E$
in the $D3/Dp$ system with $d=2+1$ dimensional intersection. The spectrum of excitations is exhausted by the holographic zero
sound mode with the speed of sound $u_0=\frac{1}{\sqrt{2}}$, and the attenuation $\Gamma_q\simeq q^2$.}\label{fig:d3coromq}
\end{figure}

We evaluate numerically the ratio ${\cal B}_E/{\cal A}_E$ on the solution of equation \rref{fleqgen-1} with incoming-wave near horizon
behavior (\ref{E-z-1}).
In Fig. \ref{fig:d3coromq} we present numerical results for the real and imaginary parts of the  ${\cal B}_E/{\cal A}_E$ for different
values of $\omega,\,q$ in the case of $b=0$. The holographic zero sound corresponds to the peak in the spectral density.



\subsection{Fluctuations of the transverse component of the gauge field}

In this subsection we will compute numerically the
holographic two-point function for the transverse current $\langle J^1(x)J^1(y)\rangle$. Let us put $b=0$.

In the near horizon regime $z\rightarrow\infty$ the bulk solution, corresponding
to the retarded current-current propagator in the dual field theory, takes the incoming-wave form
\beq
a_1=C\frac{e^{i\omega z}}{z}\,,\label{a1incwave}
\eeq
and in the vicinity of the boundary, the equation of motion becomes
\beq
a_1''-(q^2-\omega ^2)a_1=0\,,\label{a1nbound}
\eeq
with a general solution being a combination of $F_I$ and $F_{II}$ \rref{FI}, \rref{FII},
\beq
a_1={\cal A}_aF_I+{\cal B}_aF_{II}\,.
\eeq

\begin{figure}[htp]
\includegraphics[width=80mm]{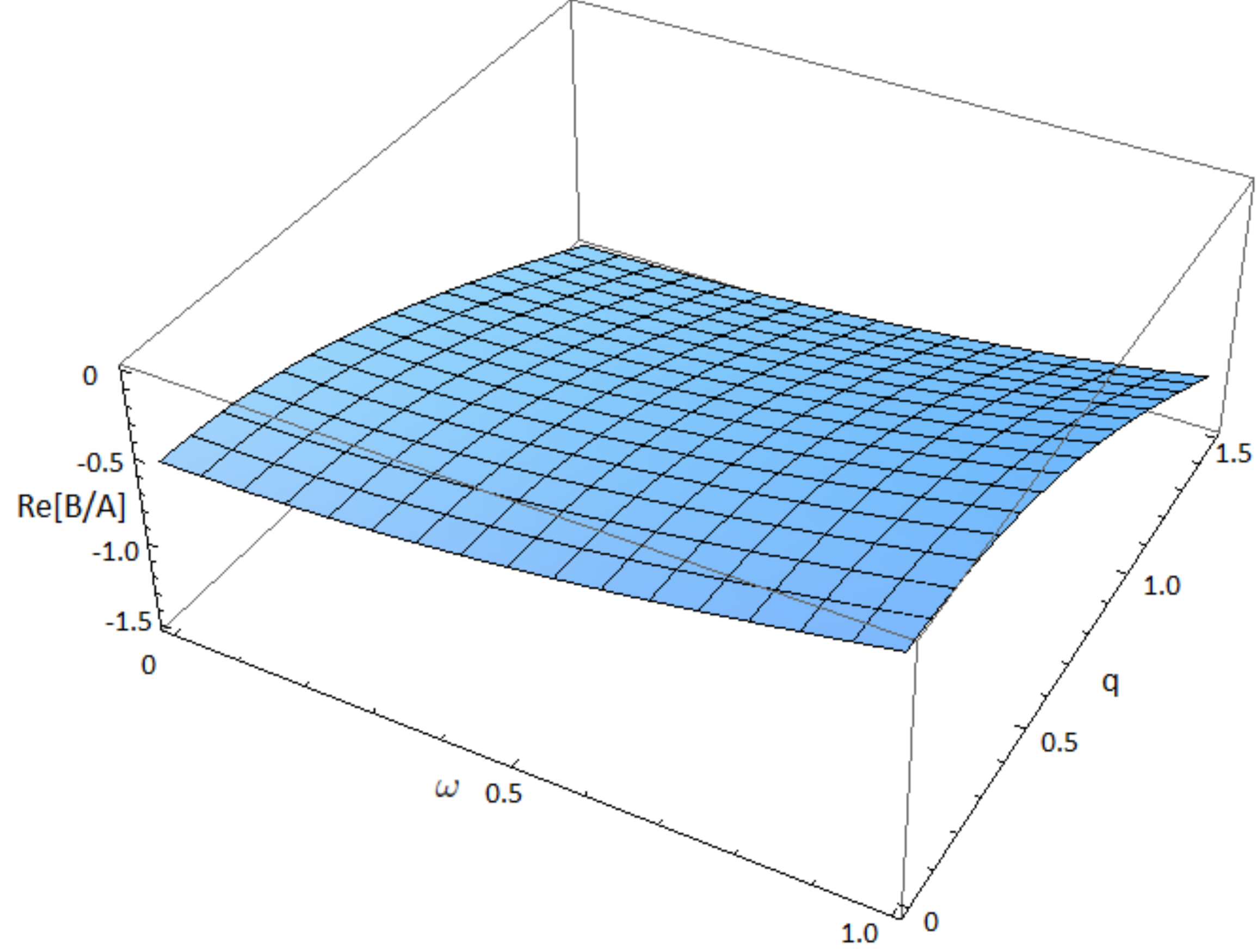}
\includegraphics[width=80mm]{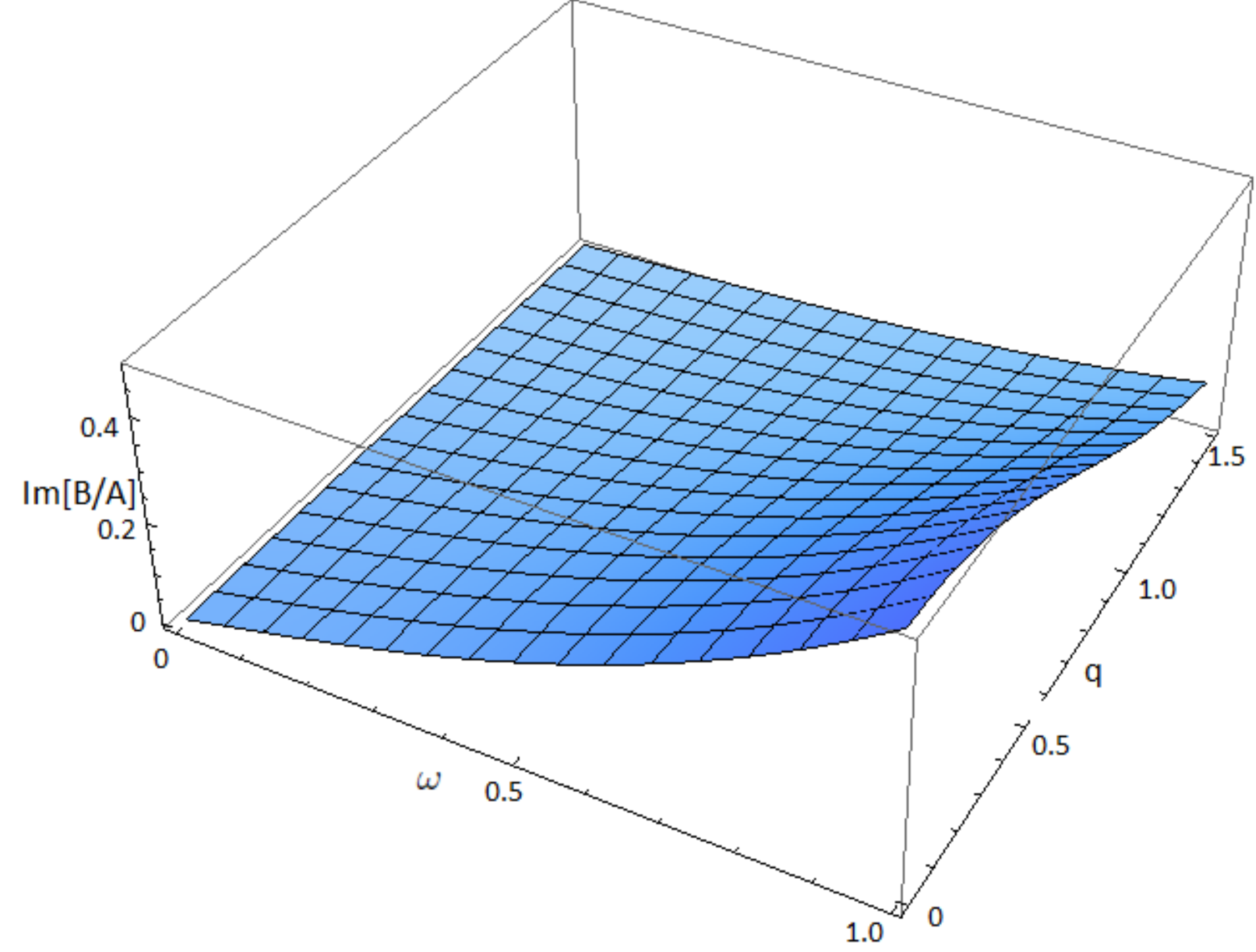}
\caption{Real and imaginary parts of the correlation function $\langle J^1(-q)J^1(q)\rangle$
in the $D3/D7$ system with $d=2+1$ dimensional intersection. No non-trivial collective
excitation modes are observed. For small frequencies
and momenta $\omega,\,q\ll 1$, the imaginary part of the correlation function behaves as $\text{Im}[\langle J^1(-q)J^1(q)\rangle]\sim\omega$,
independently of a particular value of $q$.}\label{fig:d3omqj1}
\end{figure}

The results of numerical evaluations of the holographic two-point function $\langle J^1(q)J^1(-q)\rangle=\frac{{\cal B}_a}{{\cal A}_a}$
are presented in Fig. \ref{fig:d3omqj1}.
We see that it does not reveal any structure.

\subsection{Non-vanishing magnetic field}

In the case of $b\ne 0$ fluctuations of the longitudinal $E(x^0,\,x^2,z)$ and transverse $a_1(x^0,\,x^2,z)$
components of the gauge field are no longer decoupled\footnote{We thank R. Davison for pointing this out to us.}. They are described by the action \rref{act-3},
which can be written as
\begin{align}
S&=\int dz\left(\left(-({\cal G}_EE')'+{\cal U}_EE-\frac{1}{2}({\cal C}^{(1)})'a_1\right)E+\right.\notag\\
&\left.+\left(-({\cal G}_aa_1')'+{\cal U}_aa_1-\frac{1}{2}({\cal C}^{(1)})'E\right)a_1\right)+\notag\\
&+[{\cal G}_EEE'+{\cal G}_aa_1a_1'+{\cal C}^{(1)}Ea_1]_{z=0}^{z=\infty}\,.
\end{align}
The first two lines vanish on shell. In the last line the cross term does not contribute to the variation of
the on-shell action by the boundary $z=0$ values of the fields $E$ and $a_1$, because
\beq
{\cal C}^{(1)}|_{z=0}=0\,.
\eeq
The on-shell action is then given by the boundary term
\beq
S_{on-shell}\simeq\lim_{\varepsilon\rightarrow 0}\int d\omega dq
\left(\frac{1}{q^2-\omega^2} E E'+ a_1 a_1'\right)_{z=\varepsilon}\,.\label{bonshellact}
\eeq

Near the boundary the solutions to equations of motion are given by
\beq
E={\cal A}_EF_I+ {\cal B}_EF_{II}\,,\quad\quad
a_1= {\cal A}_aF_I+ {\cal B}_aF_{II}\,,\label{nbbn01}
\eeq
where $F_{I,\,II}$ are defined by \rref{FI}, \rref{FII}.

To compute current-current two-point function numerically, we follow \cite{Kaminski:2009dh},
where general system of coupled equations in the bulk is studied. For arbitrary two independent solutions $\Phi _{(1)},\;\Phi _{(2)}$
of the coupled system of fluctuation equations \rref{fleqgen-1}, \rref{fleqa1bnon0},
we define the matrix $H =\left(\Phi _{(1)},\;\Phi _{(2)}\right)$. Near the boundary it is expanded as
\beq
H ={\cal A}F_{I}+{\cal B}F_{II}\,.\label{Hnhdef}
\eeq
On-shell action \rref{bonshellact} may be rewritten as
\beq
S_{on-shell}\simeq \int d\omega dq\, \Phi^T\,M\, \Phi'\,,
\eeq
where
\beq
M=\left({\frac{1}{q^2-\omega^2}\atop 0}\;{0\atop 1}\right)\,.
\eeq
The matrix of correlation functions is then given by (see eq. (2.34) in \cite{Kaminski:2009dh})
\beq
G\simeq M{\cal B}{\cal A}^{-1}\,.\label{GmatdefK}
\eeq
In such a form the current-current correlation matrix $G$ is explicitly independent of a linear
change of fields
\beq
\Phi _{(1)}\rightarrow r_1\Phi_{(1)}+r_2\Phi_{(2)}\,,\;\;\Phi _{(2)}\rightarrow r_3\Phi_{(1)}+r_4\Phi_{(2)}\quad\Rightarrow\quad H\rightarrow HR\,,
\eeq 
where $R=\left({{r_1\atop r_2}\;{r_3\atop r_4}}\right)$ is some arbitrary non-degenerate matrix.
If $\Phi _{(1),(2)}=\left({E^{(1),(2)}\atop a_1^{(1),(2)}}\right)$ are some arbitrary independent solutions, then
due to \rref{Hnhdef} we get
\beq
{\cal A}=\left({{\cal A}_E^{(1)}\atop {\cal A}_a^{(1)}}\;{{\cal A}_E^{(2)}\atop {\cal A}_a^{(2)}}\right)\,,\quad\quad
{\cal B}=\left({{\cal B}_E^{(1)}\atop {\cal B}_a^{(1)}}\;{{\cal B}_E^{(2)}\atop {\cal B}_a^{(2)}}\right)\,,
\eeq
and therefore using \rref{GmatdefK} we obtain
\beq
G\simeq \frac{1}{{\cal A}_E^{(1)}{\cal A}_a^{(2)}-{\cal A}_a^{(1)}{\cal A}_E^{(2)}}
\left({\frac{{\cal B}_E^{(1)}{\cal A}_a^{(2)}-{\cal B}_E^{(2)}{\cal A}_a^{(1)}}{q^2-\omega^2}
\atop {\cal B}_a^{(1)}{\cal A}_a^{(2)}-{\cal B}_a^{(2)}{\cal A}_a^{(1)}}\;
{\frac{{\cal B}_E^{(2)}{\cal A}_E^{(1)}-{\cal B}_E^{(1)}{\cal A}_E^{(2)}}{q^2-\omega^2}
\atop {\cal A}_E^{(1)}{\cal B}_a^{(2)}-{\cal B}_a^{(1)}{\cal A}_E^{(2)}}\right)\,,\label{Gcormatrix}
\eeq

Near-horizon solutions are given by \rref{nhsol1}, \rref{nhsol2}, which we can write as a linear combination of two independent solutions
\beq
\tilde \Phi _{(1)}=\left({(1-i\omega z)e^{i\omega z}\atop (1-i\sqrt{\omega ^2-b^2q^2}z)e^{i\sqrt{\omega ^2-b^2q^2} z}}\right)\,,\quad
\tilde\Phi _{(2)}=\left({(1-i\omega z)e^{i\omega z}\atop -(1-i\sqrt{\omega ^2-b^2q^2}z)e^{i\sqrt{\omega ^2-b^2q^2} z}}\right)\label{Phinh}
\eeq
Arbitrary near-horizon behavior, with the most general form (up to simultaneous rescaling of all fields by the same factor) may
therefore be written as a linear combination of these two solutions,
\beq
\Phi =\left({(1-i\omega z)e^{i\omega z}\atop c(1-i\sqrt{\omega ^2-b^2q^2}z)e^{i\sqrt{\omega ^2-b^2q^2} z}}\right)=\frac{1+c}{2}\tilde\Phi _{(1)}+\frac{1-c}{2}\tilde\Phi _{(2)}\,.
\eeq
On the other hand, fluctuation equations may be rewritten as
\beq
\Omega _1\Phi ''+\Omega _2\Phi '+\Omega \Phi=0\,,
\eeq
with matrices $\Omega _{1,2,3}$, being determined from  \rref{fleqgen-1}, \rref{fleqa1bnon0}. Therefore linear combination of
near-horizon solutions \rref{Phinh} results in the same linear combination of the solutions near the boundary.
Recall that the matrix correlation function \rref{Gcormatrix} is the same for any such a non-degenerate linear combination.

We therefore fix two arbitrary
near-horizon conditions, say \rref{Phinh}, determine corresponding coefficients ${\cal A}_E^{(1),(2)},\;{\cal A}_a^{(1),(2)}$ and
${\cal B}_E^{(1),(2)},\;{\cal B}_a^{(1),(2)}$
by integrating numerically fluctuation equations \rref{fleqgen-1}, \rref{fleqa1bnon0} up to the boundary and matching corresponding solutions
with \rref{nbbn01},
and compute the correlation matrix \rref{Gcormatrix}. Each of the four components of the correlation matrix shows
a gapped zero sound mode.

\begin{figure}[htp]
\includegraphics[width=80mm]{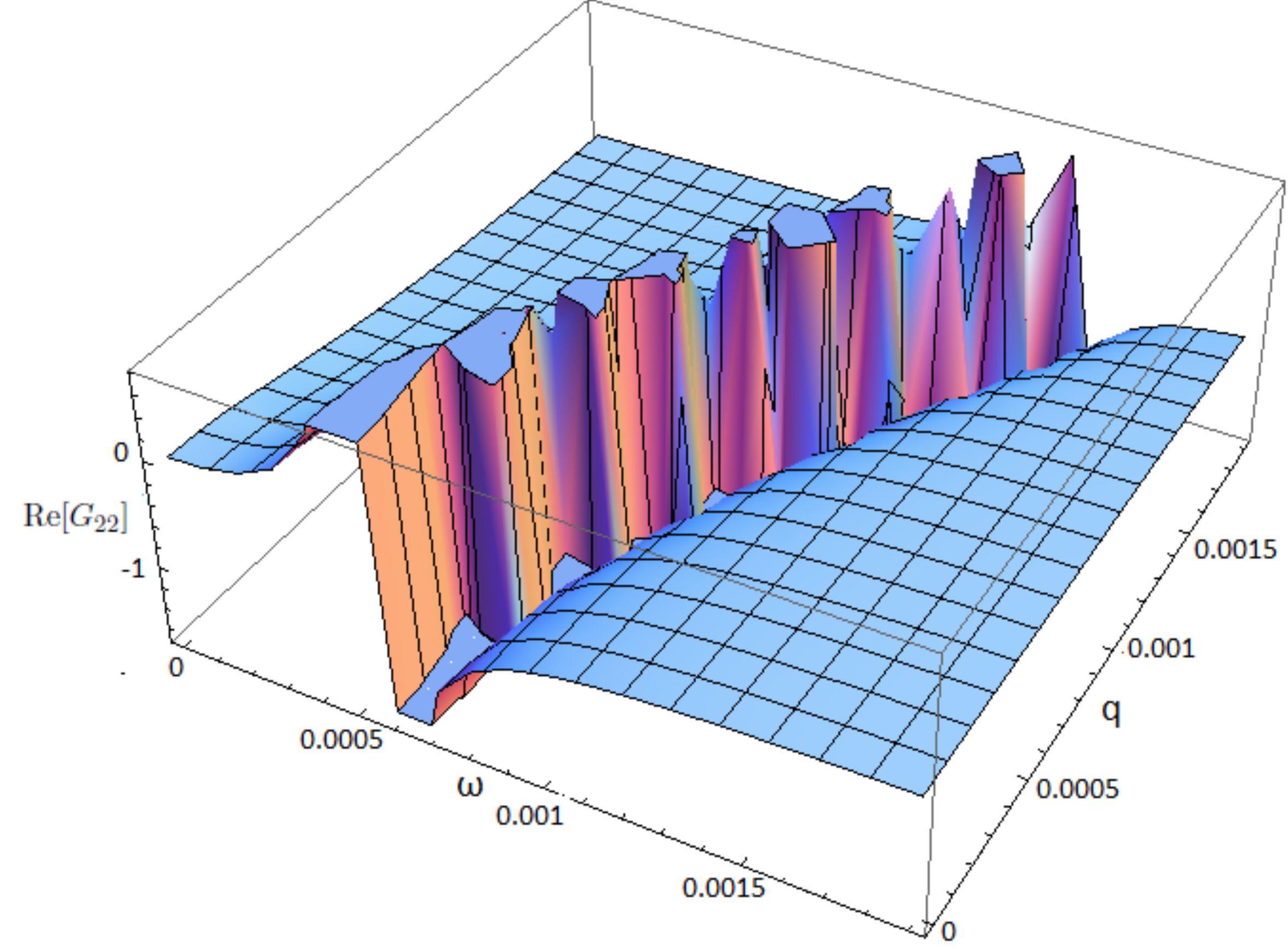}
\includegraphics[width=80mm]{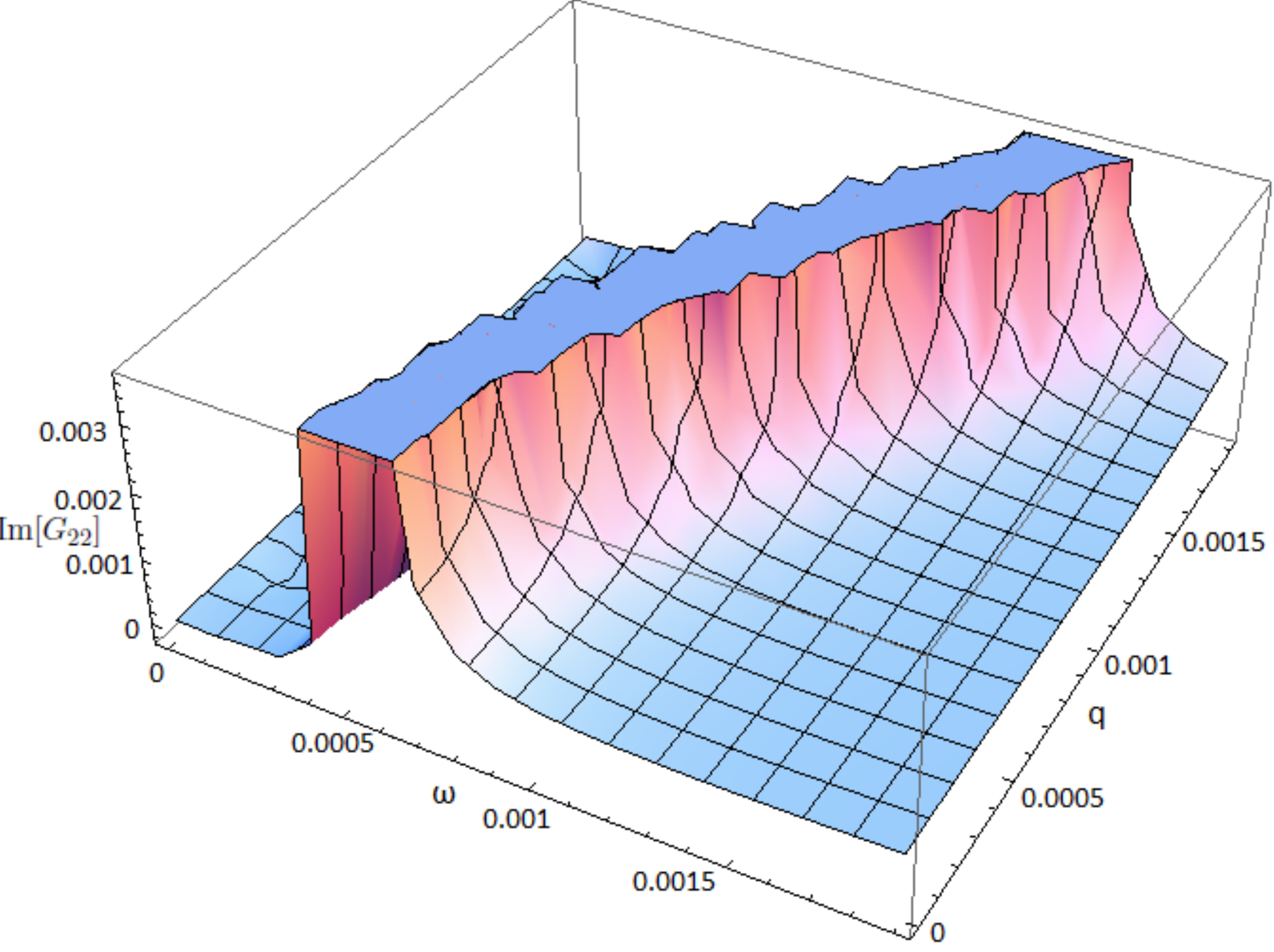}
\caption{Real and imaginary parts of the component $G_{22}$ of the correlation matrix \rref{GmatdefK}
in the $D3/D7$ system with $d=2+1$ dimensional intersection, for the magnetic field $b=0.001$. The spectrum of excitations
is exhausted by a gapped zero sound mode, with the value of the gap $\omega _c\sim b$.}\label{fig:d3omqgap}
\end{figure}

In figure \ref{fig:d3omqgap} we plot the real and imaginary parts of the $G_{22}$ component, for $b=0.001$.
We see the gapped zero sound mode, with the gap which scales as $\omega _c\sim b$, in agreement with analytic
result $\omega _c=\frac{b}{K(1/2)}$ of the previous section.


\section{Discussion}
In this paper we have studied current-current two-point functions at strong coupling.
We have considered the current-current correlators at finite momenta, but
did not observe any nontrivial structure in the spectral function, other than the
zero sound\footnote{Note that our models are different from those studied in
\cite{arXiv:0903.2477, Cubrovic:2009ye}, where poles at finite momenta were observed
in the holographic two-point functions of operators with nonvanishing charge under global $U(1)$.}.

It is instructive to compare the holographic density-density correlator with the form expected
from  the random phase approximation and reviewed in Section II.
Within RPA the zero sound mode presents itself as a smeared delta-function like peak in Fig. 1,
the Lindhard particle-hole continuum starts at $\q \simeq \w/ \upsilon_F$ and sharply ends
at $\q\simeq 2 \q_F$.
The absence of the Lindhard continuum in the holographic computations can be
explained by parametrically large values of the Landau parameters.
The key point is eq. \rref{Land-zero-sound}  which implies that since the zero
sound  velocity that we observe is $\OO(1)$, the value of Fermi velocity 
scales like $\upsilon_F\sim 1/\sqrt{F_0 F_1}$.
The regime of validity of our calculations is limited to $\w\sim \q$, and
therefore the Lindhard continuum cannot be observed for parametrically large values
of the Landau parameters.
In the following we offer some speculations on how such a scenario can play out.

We can argue that the Fermi velocity  is parametrically small.
Recall that $q\simeq \q \sqrt{\lambda}/\mu$.
Hence, eq. \rref{zeosdis} implies that the zero sound attenuation is $\alpha \sim \w^2 \sqrt{\lambda}/\mu$.
According to \cite{BaymPethick} this can be expressed in terms of the quasi-particle 
lifetime as 
\beq
\label{zsatten}
     \alpha\simeq \frac{1}{\tau}  \, \frac{m^*}{\mu}  \upsilon_F^2 F_2^2\sim \frac{\w^2}{\mu} F_0^2 F_2^2
 \eeq
To derive the second approximate equality we used the Fermi liquid estimate $1/\tau\sim \w^2 m^* {F_0}^2/\q_F^2 \sim \w^2 F_0^2/E_F^*$.
Eq. \rref{zsatten} implies that the Landau parameters are indeed parametrically large, $F_0^2 F_2^2\sim \sqrt{\lambda}$.

We have  analyzed the system in the presence of magnetic field and observed 
a gap in the excitation spectrum $\w_c$.
We derived a scaling relation $\w_c\simeq B/\mu$.
Note  that the gap in the spectrum of non relativistic fermions scales linearly with $B$,
while the relativistic fermions obey $\sqrt{B}$ scaling;
Kohn's theorem implies that the gap in the spectrum of excitations is not 
changed when the pairwise interaction is turned on.
In our setup charged fermions interact and can exchange momentum with $\NN=4$ SYM degrees of freedom;
the linear scaling of the gap with the magnetic field is consistent with the assumption that 
the effective degrees of freedom have an effective mass $m^*\simeq \mu$.
[According to eq. \rref{meffmu} this implies that $F_1=\OO (1)$; a  scenario consistent with the discussion 
above may involve a parametrically large $F_0\sim \lambda^{1/4}$  but finite $F_n, \,n>0$.]
We do not quite understand the mechanism of  dynamical mass generation
at finite density -- it is clearly very different from dynamical mass generation in a strongly
interacting fermion system at zero density\footnote{The holographic dual of the latter involves repulsion
of the probe brane from the bulk of of the AdS space; see e.g. \cite{Kutasov:2011fr} for a recent discussion.}.

We already emphasized that a priori the very existence of zero sound is nontrivial, 
given the interaction of the charged matter with the uncharged superconformal degrees of freedom.
It would be interesting to make this picture more precise and to see whether there is any relation to 
 the recent
studies of fermions in magnetic fields in the context of holography \cite{  Albash:2009wz,Albash:2010yr,Gubankova:2010rc,Bolognesi:2011un,Li:2012ib}.
It would also be interesting to compare our results with the correlators computed in the
charged magnetic brane background \cite{D'Hoker:2011xw}.

At this point it is worth recalling the relation between the charge density and the value
of the chemical potential, given by \rref{much-d}.
As usual, the value of the charge density is proportional to $\hat d$, $\rho\simeq N_c \lambda^{(p-5)/4}  \hat d$ and the proportionality
coefficient strongly depends on the dimensionality of the probe brane.
The incompressibility $\p\hat d/\p\mu$ is a smooth non-vanishing function of $\mu,b$.
This implies that we cannot rule out the existence of gapless modes in our system\footnote{We thank D. Son for pointing this out to us.}.
Indeed, the analysis that led to the existence of the zero sound implicitly assumed $\omega\sim q\sim b$,
and can be shown to break down for $|\omega|<b q$.
We leave the search for gapless quasi normal modes for future work.
Let us also note that a smooth compressibility is not compatible with the existence of
Landau levels for the effective fermions.


Appendix  is devoted to the subject of higher derivative corrections to the DBI action for the probe Dp brane.
The possibility of breakdown of the DBI description in the extreme infrared (very close to the horizon) was pointed out in \cite{Hartnoll:2009ns}.
Two possible causes were identified in the presence of the electric flux on the brane: strong back-reaction 
 and vanishing of the effective string tension.
The strength of back-reaction from the flavor branes is governed by the ratio $N_f/N _c$.
We did not investigate $1/N_c$ corrections in the paper, although it is a very interesting
problem.
Instead, we explored the effects of
the breakdown of the DBI description due to the vanishing of the effective string tension near the horizon.
The strength of this effect is controlled by an inverse power of 't~Hooft coupling.
In principle, such effects can be described by going to higher orders in the $\alpha'$ expansion
of the effective action for open strings, which corresponds to adding higher derivative terms to the DBI Lagrangian.
Unfortunately we are not aware of the precise structure of higher derivative corrections to DBI
in the presence of the worldvolume electric field.
However we were able to model this situation by writing generic higher derivative terms which become important near
the horizon and completely change the effective metric for fluctuations there.

The effect of such terms is confined to a very small region (which scales as an inverse power of 't~Hooft coupling in suitable units); outside of 
this region the second order differential equations derived from the DBI are applicable.
In principle, one can solve the higher order fluctuation equation outwards from the horizon, and then feed
the resulting solution into the second order equation.
From the point of view of the latter, this amounts to modifying the boundary conditions: an outgoing wave
(with a small coefficient) is added to the incoming wave near the horizon.
We verify that this does not introduce any qualitative new features in the two-point functions.

\section*{Acknowledgements}
We thank R. Davison, J. Erdmenger, B. Galilo,  S. Hartnoll, J. Jottar,  P. Kovtun, M. Kulaxizi, D. Kutasov,  S. Mukhin, A. O'Bannon, J. Polchinski,  G. Policastro, S. Sachdev, K. Schalm, J. Shock, E. Silverstein, D. Son and A.Starinets
for useful discussions and comments on the manuscript. A.P. and J.Z. are grateful to the organizers of the AdS/CMT program at KITP, Santa Barbara,
where part of this work was completed.
A.P. is grateful to the University of Chicago and University of Washington, where parts of this work were completed, for hospitality.
This work was supported in part by the VIDI innovative research grant from NWO.

\section*{\textbf{Appendix}: Higher-derivative corrections to $L_{DBI}(a_1)$}
The DBI description might break down
in the near-horizon region \cite{Hartnoll:2009ns}, and therefore higher derivative
corrections become essential in that region.
Consider higher derivative correction to the DBI Lagrangian of the form \cite{Myers:2009ij, Myers:2010pk}
\beq
\frac{\epsilon}{2}\sqrt{-g}g^{\mu\lambda}g^{\nu\rho}g^{\alpha\beta}(\nabla _\alpha F_{\lambda\rho})(\nabla _\beta F_{\mu\nu})\,,\label{corr-2}
\eeq
Alternatively, using the Bianchi identity, one may rewrite it as $\tilde\epsilon\sqrt{-g}g_{\mu\nu}\nabla _\lambda F^{\lambda\mu}\nabla _\sigma F^{\sigma\nu}$.
Here $\epsilon\sim\ell _s^2\sim\frac{1}{\sqrt{\lambda}}$.

In this section we put $L=1$.
The induced $AdS_4\times S^4$ metric on the trivially embedded Dp brane world-volume then takes the form
\beq
ds^2=\rho ^2(-(dx^0)^2+(dx^1)^2+(dx^2)^2)+\frac{d\rho ^2}{\rho ^2}+d\Omega _4^2\,,\label{Ads-induced}
\eeq
This corresponds to non-vanishing Christoffel symbols in the AdS subspace,
\beq
\Gamma ^\rho _{\rho\rho}=-\frac{1}{\rho}\,,\quad\Gamma ^\rho _{ij}=-\rho ^3\eta _{ij}\,,\quad\Gamma ^i_{\rho j}=\frac{1}{\rho}\delta ^i_j\,,
\eeq
where $\eta _{00}=-1,\,\eta _{11}=\eta _{22}=1$.
We fix the background value of $A_0'(\rho)$ \rref{A0backgr} (with $\bar B=0$)
and study the dynamics of the fluctuation field $a_1(\rho,x^0,x^2)$. Consequently, the non-vanishing components of the field strength tensor
covariant derivatives
\beq
\nabla _\alpha F_{\mu\nu}=\partial _\alpha F_{\mu\nu}-\Gamma ^\tau _{\alpha\mu}F_{\tau\nu}-\Gamma^\tau _{\alpha\nu}F_{\mu\tau}\,
\eeq
are given by
\beq
\nabla _1F_{0\rho}=-\frac{1}{\rho}F_{01}\,,\quad \nabla _{\rho}F_{0\rho}=\p_\rho F_{0\rho}\,,\label{df1}
\eeq
\beq
\n _1F_{01}=\rho ^3F_{0\rho}\,,\quad \n_2F_{01}=\p_2F_{01}\,,\quad\n _0F_{01}=\p _0F_{01}-\rho ^3F_{\rho 1}\,,\quad
\n_\rho F_{01}=\p _\rho F_{01}-\frac{2}{\rho}F_{01}\,,\label{df2}
\eeq
\beq
\n_2F_{12}=\p_2F_{12}+\rho ^3F_{1\rho}\,,\quad \n_0F_{12}=\p_0F_{12}\,,\quad \n_\rho F_{12}=\p_\rho F_{12}-\frac{2}{\rho}F_{12}\,,\label{df3}
\eeq
\beq
\n_2F_{\rho 1}=\p _2F_{\rho 1}-\frac{1}{\rho}F_{21}\,,\quad\n _0F_{\rho 1}=\p _0F_{\rho 1}-\frac{1}{\rho}F_{01}\,,\label{df4}
\quad\n_\rho F_{\rho 1}=\p_\rho F_{\rho 1}\,.
\eeq

Let us now substitute the quantities (\ref{df1})-(\ref{df4})
into the Lagrangian \rref{corr-2}, which becomes in momentum representation,
\bea
\Delta L&=&\epsilon [-\rho ^2(\p_\rho A_0)^2-\rho ^4(\p_\rho ^2A_0)^2+\frac{1}{\rho ^2}\left(\frac{(q^2-\omega ^2)^2}{\rho^2}
+5q^2-6\omega ^2\right)a_1^2+2(\rho^2+q^2-\omega ^2)(\p_\rho a_1)^2 \notag\\
&+&\rho^4 (\p_\rho^2 a_1)^2+\frac{4(\omega^2-q^2)}{\rho}a_1\,\p_\rho a_1]\,.\label{DfDf}
\eea
To obtain the corrected equation of motion of the background field $\p_\rho A_0$,
we put the $a_1$ fluctuations to zero and write the
total, DBI + corrections, Lagrangian as
\beq
L=\rho ^2\sqrt{1-(\p_\rho A_0)^2}-\epsilon[\rho ^2(\p_\rho A_0)^2+\rho ^2(\p_\rho ^2 A_0)^2]\,.
\eeq
The corresponding equation of motion
\beq
\frac{\rho ^2\p_\rho A_0}{\sqrt{1-(\p_\rho A_0)^2}}+2\epsilon[\rho ^2\,\p_\rho A_0-\p_\rho\left(\rho ^4\,\p_\rho^2 A_0\right)]=\hat d^2
\eeq
is solved to first order in $\epsilon$ by
\beq
\p_\rho A_0=\frac{\hat d^2}{\sqrt{\rho ^4+\hat d^4}}+\delta\,\p_\rho A_0\,,
\eeq
where we have denoted the correction to the background as
\beq
\delta\,\p_\rho A_0 =-\frac{2\hat d^2\epsilon\rho ^6(\hat d^8+16\hat d^4\rho ^4+3\rho ^8)}{(\rho ^4+\hat d^4)^4}\,.\label{deltaA0}
\eeq
Note that (\ref{deltaA0}) approaches zero as $\OO(\rho ^6)$, near the horizon $\rho =0$.
Therefore the correction to the behavior of the background potential $\p_\rho A_0$ does not substantially affect the near-horizon physics.
Using the $z=1/\rho$ radial coordinate, and considering the near horizon limit $\omega z\gg 1$, we obtain from \rref{DfDf} the
correction to the near-horizon DBI Lagrangian
\beq
\Delta L=\epsilon\left((q^2-\omega ^2)z^2(2a_1^{\prime 2}+(q^2-\omega ^2)a_1^2)+(2a_1'+za_1'')^2\right)\,.\label{za1-corr}
\eeq
This is to be added to the quadratic DBI near-horizon Lagrangian, 
\beq
L_{DBI}=z^2(a_1^{\prime 2}-\omega ^2a_1^2)\,.\label{za1orig}
\eeq
As a result we obtain the following near-horizon Lagrangian:
\beq
L=\left(\left(1+2\epsilon (q^2-\omega ^2)\right)z^2+2\epsilon\right)a_1^{\prime 2}+\left(-\omega ^2+\epsilon (q^2-\omega ^2)^2\right)z^2a_1^2+
2\epsilon (za_1^{\prime 2})'+\epsilon z^2(a_1'')^ 2\,.
\eeq
Up to a total derivative term\footnote{Corresponding boundary terms $2\epsilon za_1^{\prime 2}$, evaluated on non-perturbed solution $e^{i\omega z}/z$,
vanish when $z\gg 1$.} and $\OO(\epsilon)$ modification of the DBI behavior,
this Lagrangian therefore may be rewritten as
\beq
L=z^2(a_1^{\prime 2}-\omega ^2a_1^2)+\epsilon z^2(a_1'') ^2\,,
\eeq
with associated equation of motion
\beq
a_1''+\frac{2}{z}a_1'+\omega ^2a-\epsilon \left(a_1''''+\frac{4}{z}a_1'''\right)=0\,.
\eeq
To estimate the relative significance of the correction and DBI terms, let us compare terms $a_1''$ and $\epsilon \left(a_1''''+\frac{4}{z}a_1'''\right)$, when evaluated on the non-perturbed near-horizon solution $e^{i\omega z}/z$:
\beq
a_1''\simeq\frac{1+(\omega z)^2}{z^3}\,,\quad\quad\epsilon \left(a_1''''+\frac{4}{z}a_1'''\right)\simeq\epsilon\frac{\omega ^4}{z}\,.
\eeq
We observe that this correction is negligible.

\begin{figure}[htp]
\includegraphics[width=100mm]{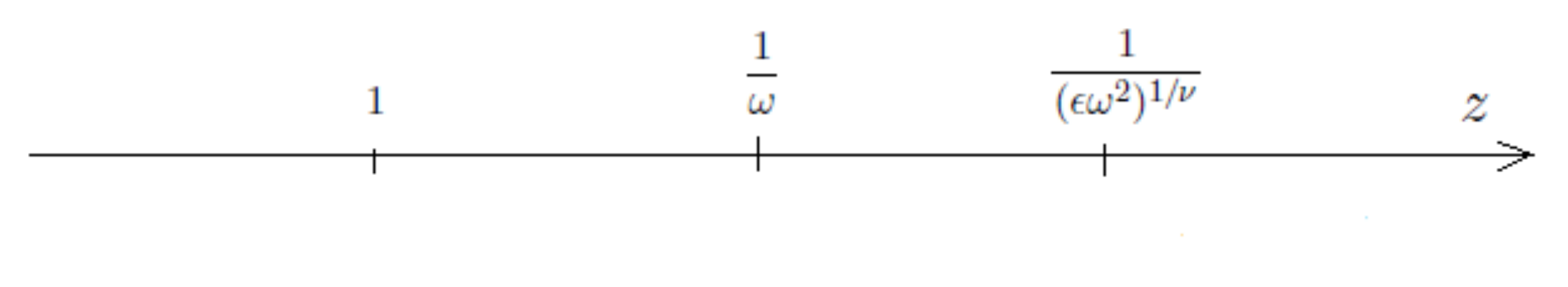}
\caption{Hierarchy of scales in the near-horizon region for $0<\nu\leq 2$.}\label{zlinel1}
\end{figure}

\begin{figure}[htp]
\includegraphics[width=100mm]{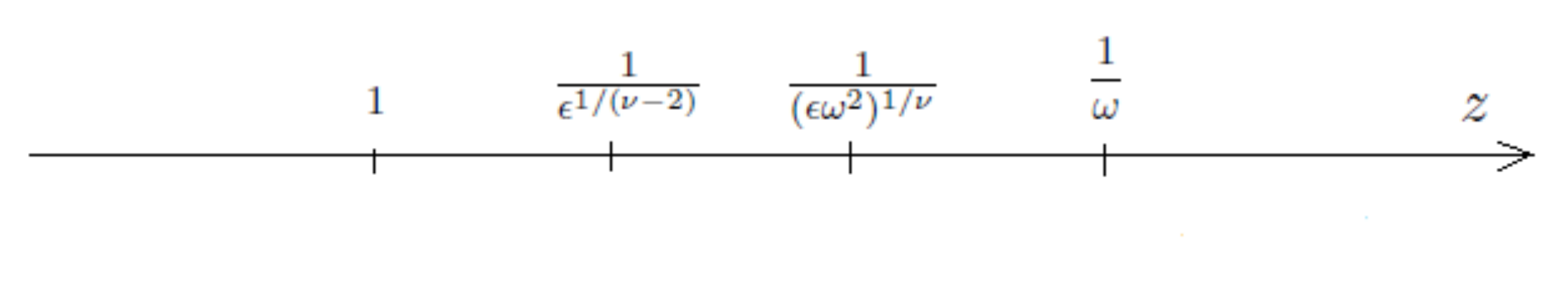}
\caption{Hierarchy of scales in the near-horizon region for $\nu >2$ and $\omega<\epsilon ^{1/(\nu-2)}$.}\label{zlinel2}
\end{figure}

Unfortunately we are not aware of the exact form of the higher derivative corrections to 
the DBI action in the presence of the electric field on the world-volume of the probe Dp brane.
In the following we will simply assume a particular expression for the higher derivative
corrections to the Lagrangian for the transverse fluctuations:
\beq
L=z^2(a_1^{\prime 2}-\omega ^2a_1^2)+\epsilon z^{2+\nu}(a_1'')^2\,,
\eeq
with $\nu >0$. 
To estimate the significance of the correction term we need  to compare
contributions to the equation of motion from the terms  $\OO(1)$ 
\beq
a_1''\simeq\frac{1+(\omega z)^2}{z^3}
\eeq
and  $\OO(\epsilon)$ 
\beq
\epsilon \left(z^\nu a_1''''+2(\nu +2)z^{\nu -1}a_1'''+(\nu +1)(\nu +2)z^{\nu -2} a_1''\right)\simeq\epsilon z^{\nu -5}(1+(\omega z)^4)\,.
\eeq
Therefore, if $0<\nu\leq 2$, the correction becomes significant when $z\gg\frac{1}{(\epsilon\omega ^2)^{1/\nu}}$
(see the hierarchy of scales in Fig. \ref{zlinel1}).
If $\nu >2$, considering modes with sufficiently low frequency  $\omega <\epsilon ^{1/(\nu-2)}$ the correction becomes
significant when $z\gg\epsilon ^{1/(2-\nu)}$ (see Fig. \ref{zlinel2}). Finally, if $\nu >2$ and $\omega >\epsilon ^{1/(\nu-2)}$,
the Fig. \ref{zlinel1} is applicable, and the correction is significant when $z\gg\frac{1}{(\epsilon\omega ^2)^{1/\nu}}$.
Hence, in the region   $z\ll\frac{1}{(\epsilon\omega ^2)^{1/\nu}}$ the DBI description is valid,
provided  that $0<\nu \leq 2$ or $\nu >2,\;\omega>\epsilon ^{1/(\nu-2)}$.
The DBI description is valid in the region $z\ll\epsilon ^{1/(2-\nu)}$ for $\nu>2,\;\omega <\epsilon ^{1/(\nu-2)}$.

The behavior of $a_1$ in the  limit $z\gg1$ where the DBI description is valid, is different from the incoming-wave \rref{a1incwave}:
it has a qualitative form of ``incoming wave" + $\OO(\epsilon)$ ``outgoing wave".
It is worth noting that the effect of higher derivative corrections
on the current-current correlation function is essentially the same as an effect of non-zero $b$-field. We verified that such a modification
does not lead to any nontrivial structure in the spectral density.

\end{document}